\documentclass[prd,a4]{revtex4}
\usepackage[english]{babel}
\usepackage[utf8]{inputenc}
\usepackage{amsmath,amssymb}
\usepackage[dvips]{graphicx}
\usepackage{ifthen,array}
\usepackage[sort&compress]{natbib}
\usepackage{amsfonts}
\usepackage{graphicx}
\usepackage{psfrag}
\usepackage{pstricks}
\usepackage{moreverb}
\DeclareMathAlphabet{\mathsc}{OT1}{cmr}{m}{sc}
\allowdisplaybreaks

\def\vb#1{\vbox to #1 pt{}}

\newcommand{\AddrLisb}{%
 Departamento de F\'\i sica and CFTP, Instituto Superior T\'ecnico\\
          Avenida Rovisco Pais 1, $\:\:$ 1049-001 Lisboa, Portugal }

\begin{document}

\vspace*{2cm} \title{The need for the Higgs boson in the Standard Model}

\author{J.~C.~Rom\~ao}
\email{jorge.romao@tecnico.ulisboa.pt}\affiliation{\AddrLisb}

\begin{abstract}
We review the role of the Higgs boson in preserving
unitarity of the scattering amplitudes in the Standard Model
(\texttt{SM}). We will look at the processes 
$\nu_e + \overline{\nu}_e \rightarrow W^-_L +W^+_L $, $ W^-_L + W^+_L
\rightarrow W^-_L +W^+_L $ and $e^- + e^+ \rightarrow W^-_L +W^+_L $ for
longitudinally polarized gauge bosons. Special emphasis will be put in
using algebraic methods to evaluate the amplitudes and cross
sections. This note is based on Lectures given at the IDPASC
  Schools at Udine (2012) and Braga (2014).
\end{abstract}

\maketitle

\section{Introduction}

As we will be talking about cancellations, the conventions for the
vertices of the \texttt{SM} are very important. So we will collect here all the
necessary couplings for our purposes. We will also discuss the
polarization vectors of the gauge bosons, particularly the
longitudinal polarization vector and the implications of unitarity on
the growth of the amplitudes with $\sqrt{s}$.

\subsection{Gauge Boson Self-Couplings}
\begin{figure}[!htb]
\begin{tabular}{ll}
\psfrag{p}{$p$}
\psfrag{q}{$q$}
\psfrag{k}{$k$}
\psfrag{W1}{$W_{\alpha}^-$}
\psfrag{W2}{$W_{\beta}^+$}
\psfrag{A}{$A_{\mu}$}
\includegraphics[width=0.15\textwidth]{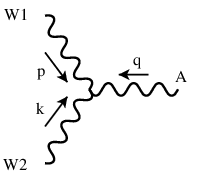}&
\begin{minipage}{0.65\linewidth}
\vspace{-30mm}
  $ -i e \left[ g_{\alpha\beta} (p - k)_\mu
 +g_{\beta\mu} (k-q)_{\alpha} + g_{\mu\alpha}
 (q-p)_{\beta}\right]\equiv  -i e \Gamma_{\alpha\beta\mu}(p,k,q)$
\end{minipage}
\\[+3mm]
\psfrag{p}{$p$}
\psfrag{q}{$q$}
\psfrag{k}{$k$}
\psfrag{W1}{$W_{\alpha}^-$}
\psfrag{W2}{$W_{\beta}^+$}
\psfrag{Z}{$Z_{\mu}$}
\includegraphics[width=0.15\textwidth]{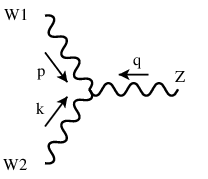}&
\begin{minipage}{0.7\linewidth}
\vspace{-30mm}
$ i g \cos\theta_W \left[ g_{\alpha\beta} (p - k)_\mu
 +g_{\beta\mu} (k-q)_{\alpha} + g_{\mu\alpha}
 (q-p)_{\beta}\right]\equiv i g  \Gamma_{\alpha\beta\mu}(p,k,q) $
\end{minipage}\\[+3mm]
\psfrag{W1}{$W_{\alpha}^-$}
\psfrag{W2}{$W_{\beta}^+$}
\psfrag{W3}{$W_{\mu}^+$}
\psfrag{W4}{$W_{\nu}^-$}
\includegraphics[width=0.15\textwidth]{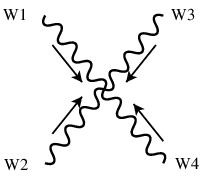}&
\begin{minipage}{0.42\linewidth}
\vspace{-30mm}
$ i g^2 \left[ 2 g_{\alpha\nu} g_{\beta\mu} -
g_{\alpha\mu} g_{\beta\nu} -g_{\alpha\beta} g_{\mu\nu} \right]$
\end{minipage}
\end{tabular}
  \caption{Gauge boson self-couplings}
  \label{fig:gaugecouplings}
\end{figure}

\noindent
where we have defined, for future convenience,
\begin{equation}
  \label{eq:20}
  \Gamma_{\alpha\beta\mu}(p,k,q)\equiv\left[ g_{\alpha\beta} (p - k)_\mu
 +g_{\beta\mu} (k-q)_{\alpha} + g_{\mu\alpha}
 (q-p)_{\beta}\right]
\end{equation}
with the convention for the momenta and charge of the particle,
given in the figure.

\subsection{Gauge Couplings to Fermions}

\begin{figure}[!ht]
  \centering
\psfrag{x1}{$\psi_{d,u}$}
\psfrag{x2}{$\psi_{u,d}$}
\psfrag{x3}{$\psi_{f}$}
\psfrag{W}{$W^{\pm}_{\mu}$}
\psfrag{Z}{$Z_{\mu}$}
\psfrag{A}{$A_{\mu}$}
\psfrag{v1}{$\displaystyle i 
\frac{g}{\sqrt{2}}\gamma_{\mu}\frac{1-\gamma_5}{2}$}
\psfrag{v2}{$\displaystyle i 
\frac{g}{\cos\theta_W}\gamma_{\mu}\left(g_V^f-g_A^f \gamma_5\right)$}
\psfrag{v3}{$\displaystyle - i e Q_f \gamma_{\mu} $}
\hskip -10mm
\includegraphics[width=160mm]{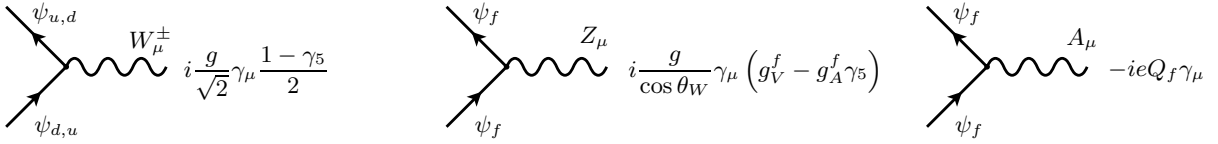}
  \caption{Gauge couplings to fermions}
  \label{fig:gaugefermions}
\end{figure}
where 
\begin{equation}
  \label{eq:4}
  g_V^f = \frac{1}{2} T_3^f -Q_f \sin^2\theta_W,\quad
  g_A^f = \frac{1}{2} T_3^f
\end{equation}

Sometimes it is more useful to write these couplings in terms of the
left and right projectors
\begin{equation}
  \label{eq:5}
  P_L=\frac{1-\gamma_5}{2},\quad   P_R=\frac{1+\gamma_5}{2},\quad
\end{equation}
We get
\begin{equation}
  \label{eq:6}
  g_V^f-g_A^f \gamma_5 =  g_L^f P_L + g_R^f P_R
\end{equation}
where
\begin{equation}
  \label{eq:6a}
   g_L^f= g_V^f+g_A^f= T_3^f -Q_f
  \sin^2\theta_W,\qquad g_R^f= g_V^f-g_A^f= -Q_f \sin^2\theta_W
\end{equation}

\subsection{Higgs Boson Couplings}

\begin{figure}[!ht]
\rput(-1,-1){
\begin{minipage}{1.0\linewidth}
\begin{tabular}{llllll}
\includegraphics[width=0.15\textwidth]{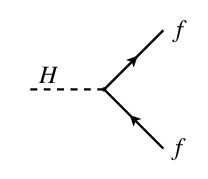}&
\put(-5,10){\mbox{$ -i \displaystyle\frac{g}{2}\, \displaystyle
\frac{m_f}{m_W}$}}&
\hskip 20mm 
\includegraphics[width=0.15\textwidth]{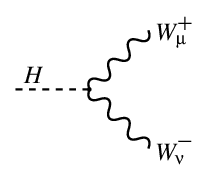}&
\put(-5,10){\mbox{$  i g\, m_W\, g_{\mu\nu}$}}&
\hskip 20mm 
\includegraphics[width=0.15\textwidth]{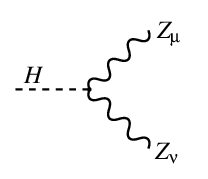}&
\put(-5,10){\mbox{$ i \displaystyle
\frac{g}{\cos\theta_W}\, m_Z\, g_{\mu\nu}$}}
\end{tabular}
\end{minipage}
}
\vspace{20mm}
  \caption{Higgs boson couplings}
  \label{fig:higgs}
\end{figure}

\subsection{Polarization Vectors}

In many problems with massive gauge bosons we do not measure their
polarization, and therefore we sum over all polarizations using the
well known result,
\begin{equation}
  \label{eq:44}
  \sum_{\lambda} \varepsilon^{\mu}(k,\lambda)
  \varepsilon^{* \nu}(k,\lambda) = - g^{\mu\nu} + \frac{k^{\mu}k^{\nu}}{M_W^2}
\end{equation}
where we used the $W$ boson as an example.  As we will be considering
the case of longitudinal polarized gauge bosons, we have to review the
expressions for the polarization vectors. Let us start with the case
of longitudinal polarization.  In the gauge boson rest frame (we
consider here the case of the $W$, but similar expressions apply also
for the case of the $Z$) where $p^\mu=(M_W,0,0,0)$, the longitudinal
polarization vector is
\begin{equation}
  \label{eq:7}
  \varepsilon^\mu_L(p)=(0,0,0,1), \quad \hbox{satisfying}\quad
 \varepsilon_L(p)\cdot \varepsilon_L(p)=-1, \varepsilon_L(p)\cdot p
 =0\ .
\end{equation}
In the frame where the gauge boson is moving with velocity $\vec \beta$
 we have
 \begin{equation}
   \label{eq:12}
  \varepsilon^\mu_L(p)= (\gamma \beta, \gamma \hat \beta)   
 \end{equation}
where, as usually, $\vec \beta= \vec p/E$,
$\gamma^{-1}=\sqrt{1-\beta^2}$ e $\hat \beta= \vec
\beta/\beta$, verifying the invariant relations 
$\varepsilon_L(p)\cdot \varepsilon_L(p)=-1$ e $\varepsilon_L(p)\cdot p
=0$. For $E\gg M_W$ we also have approximately 
\begin{equation}
  \label{eq:8}
  \varepsilon^\mu_L(p)\simeq \frac{p^\mu}{M_W}
\end{equation}
We should notice that this relation is only approximate and can not be
applied in all cases (the invariant relations are violated by terms
of the order of $M_W^2/E^2$). In fact, one can easily show that a
better approximation is
\begin{equation}
  \label{eq:9}
  \varepsilon^\mu_L(p) = \frac{p^\mu}{M_W} + \frac{1}{2\gamma^2}
  \frac{Q^\mu}{M_W} + \cdots
\end{equation}
where
\begin{equation}
  \label{eq:10}
  Q^\mu = \left( -E, \vec p\right)
\end{equation}
With this relation one can show that
\begin{equation}
  \label{eq:11}
  \varepsilon_L(p)\cdot \varepsilon_L(p) = -1 +
  \mathcal{O}\left(\frac{1}{\gamma^4}\right), \quad
   \varepsilon_L(p)\cdot p = 0 + \mathcal{O}\left(\frac{1}{\gamma^4}\right)
\end{equation}
We will see below that, depending on the case, one can use the simpler
Eq.~(\ref{eq:8}), the more correct Eq.~(\ref{eq:9}) or the exact
expression in Eq.~(\ref{eq:12}). 

\begin{figure}[!ht]
  \centering
\psfrag{x}{$x$}
\psfrag{xp}{$x'$}
\psfrag{z}{$z$}
\psfrag{zp}{$z'$}
\psfrag{t}{$\theta$}
\psfrag{y}{$y=y'$}
\psfrag{k}{$\vec k$}
  \includegraphics[width=55mm]{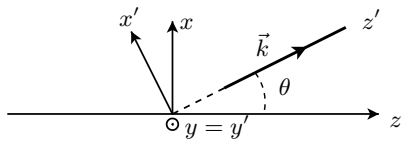}
\vspace{-5mm}
  \caption{Kinematics of the gauge boson}
  \label{fig:kinematics}
\end{figure}

Now we consider the case of transverse polarized gauge bosons. Here we
have in the frame where the gauge boson moves along the $z$ axis, that
is, $p^{\mu}=E_W(1, 0,0,\beta)$ that the two degrees of transverse
polarization can be written as
\begin{equation}
  \label{eq:45}
  \varepsilon^{\mu}_T=(0,\vec \varepsilon_{T_{1,2}})
\end{equation}
where $\vec \varepsilon_{T_{1,2}}$ lie in the plane perpendicular to the
$z$ axis. Any combination of linearly independent vectors will do, but
there are obviously two preferred choices. The first one, corresponds
to two linearly polarized perpendicular vectors,
\begin{equation}
  \label{eq:46}
  \varepsilon^{\mu}_{T_1} = (0,1,0,0),\quad \varepsilon_{T_1} = (0,0,1,0)
\end{equation}
and the other choice corresponds to helicity $\pm$,
\begin{equation}
  \label{eq:47}
  \varepsilon^{\mu}_{\pm} = (0, \frac{1}{\sqrt{2}}, \pm \frac{i}{\sqrt{2}}, 0) 
\end{equation}
satisfying the correct normalization conditions $\varepsilon(p)\cdot
\varepsilon^*(p)=-1, \varepsilon(p)\cdot p =0$. Note that for the
case of helicity vectors, as they are complex, we have to take the
complex conjugate in the normalization condition. In real situations
the gauge bosons will be moving along some direction $\vec \beta =
\vec k/E_W$ with respect to some reference frame, usually defined by
the incident particles. If we have the situation indicated in Fig.~\ref{fig:kinematics} 
then we will have for the two choices of polarization vectors,
\begin{equation}
  \label{eq:48}
  \varepsilon^{\mu}(\lambda=\pm)=(0, \frac{1}{\sqrt{2}}\cos\theta, \pm
  \frac{i}{\sqrt{2}},- \frac{1}{\sqrt{2}} \sin\theta)
\end{equation}
and 
\begin{equation}
  \label{eq:49}
  \varepsilon^{\mu}(\lambda=\vec e_x)= (0,\cos\theta,0,-\sin\theta),\quad
  \varepsilon^{\mu}(\lambda=\vec e_y)= (0,0,1,0)
\end{equation}

We leave it for the reader to verify that using the definitions of
Eqs.~(\ref{eq:12}) and (\ref{eq:48}) or (\ref{eq:49}) we can reproduce
the general result of the sum over all polarizations,
Eq.~(\ref{eq:44}). Also a more compact expression for the polarization
vector, valid for all polarizations,  can be shown to be given
by~\cite{quigg:1983}, 
\begin{equation}
  \label{eq:50}
  \varepsilon^{\mu} = \left[ \gamma (\vec \beta\cdot\vec\varepsilon),
    \vec \varepsilon + (\gamma -1)
    (\hat\beta\cdot\vec\varepsilon)\hat\beta \right]
\end{equation}
where $\vec \varepsilon$ is the polarization vector in the rest frame
of the gauge boson.

\subsection{Unitarity and Growth of the Amplitudes with $\sqrt{s}$}

The last topic we want to discuss in this introduction is the
behavior of the amplitudes with the growth of the center of mass
energy. For these $1+2\rightarrow 3+4$ processes, it can be
shown~\cite{quigg:1983}, that
the amplitudes for large values of $\sqrt{s}$ should, at most, be
constant with the energy,
\begin{equation}
  \label{eq:14}
  \lim_{\sqrt{s}\rightarrow\infty} \mathcal{M} = \hbox{constant.}
\end{equation}
 This in turn will imply that the cross sections given
 by~\cite{romao:2016itc}, 
\begin{equation}
  \label{eq:13}
  \frac{d\sigma}{d\Omega}= \frac{1}{64\pi^2\, s}\ \frac{|\vec p_{\rm 3
    CM}|}{|\vec p_{\rm 1 CM}|}\ \overline{|\mathcal{M}|}^2
\end{equation}
will decrease for values of $\sqrt{s}\gg M$, where $M$ is any mass in
the problem.

\section{The scattering $\nu_e + \overline{\nu}_e \rightarrow W^+_L  +W^-_L $}
\label{sec:nunuWLWL}

\begin{figure}[htb]
  \centering
\psfrag{n}{$\nu_e$}
\psfrag{nb}{$\overline{\nu}_e$}
\psfrag{Wp}{$W^+$}
\psfrag{Wm}{$W^-$}
\psfrag{e}{$e$}
\psfrag{g}{$\gamma$}
\psfrag{Z}{$Z$}
  \includegraphics{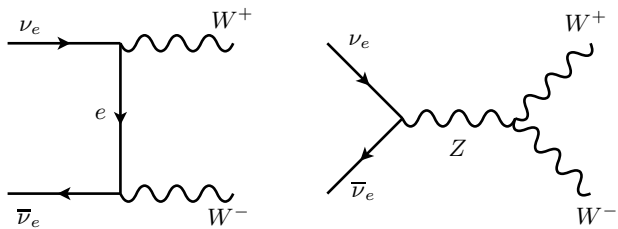}
\vspace{-3mm}
  \caption{Diagrams contributing to $\nu_e + \overline{\nu}_e \rightarrow W^+_L  +W^-_L$.}
  \label{fig:1}
\end{figure}

As a first exercise on the structure of the \texttt{SM} we will look
at the process 
\begin{equation}
  \label{eq:15}
  \nu_e(p_1) + \overline{\nu}_e(p_2) \rightarrow W^+_L(q_1)  +W^-_L(q_2) 
\end{equation}
In the \texttt{SM} this process has the tree-level diagrams shown in
Fig.~\ref{fig:1}. 

\begin{figure}[htb]
  \centering
\psfrag{t}{$\theta_{\rm CM}$}
\psfrag{p1}{$p_1$}
\psfrag{p2}{$p_2$}
\psfrag{q1}{$q_1$}
\psfrag{q2}{$q_2$}
  \includegraphics{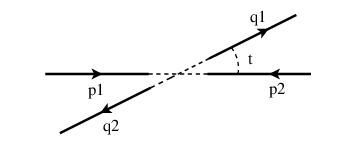}
\vspace{-3mm}
  \caption{\texttt{CM} kinematics.}
  \label{fig:CM}
\end{figure}

\noindent
The kinematics in the \texttt{CM} frame is given in Fig.~\ref{fig:CM},
where $\theta_{\rm CM}$ is the scattering angle in the \texttt{CM}
frame.
which can be written as
\begin{equation}
  \label{eq:18}
  \left\{
    \begin{array}{l}\displaystyle
      p_1=\frac{\sqrt{s}}{2} (1, 0,0,1)\\[+2mm]
      \displaystyle
      p_2=\frac{\sqrt{s}}{2} (1, 0,0,-1)\\[+2mm]
      \displaystyle
      q_1=\frac{\sqrt{s}}{2} (1, \beta\sin\theta_{\rm
        CM},0,\beta\cos\theta_{\rm CM})\\[+2mm] 
      \displaystyle
      q_2=\frac{\sqrt{s}}{2} (1, -\beta\sin\theta_{\rm
        CM},0,-\beta\cos\theta_{\rm CM})\\[+2mm] 
    \end{array}
  \right.
\end{equation}
where $\beta$ is the velocity of the $W$ in the \texttt{CM} frame,
\begin{equation}
  \label{eq:17}
  \beta=\sqrt{1 - \frac{4M_W^2}{s}}
\end{equation}
It is also important to write down the explicit expressions for the
longitudinal polarization vectors. We have
\begin{equation}
  \label{eq:19}
    \begin{array}{l}
      \displaystyle
      \varepsilon_L^{\mu}(q_1)=\frac{\sqrt{s}}{2M_W} (\beta, \sin\theta_{\rm
        CM},0,\cos\theta_{\rm CM})\\[+3mm] 
      \displaystyle
      \varepsilon_L^{\mu}(q_2)=\frac{\sqrt{s}}{2M_W} (\beta, -\sin\theta_{\rm
        CM},0,-\cos\theta_{\rm CM})\\[+3mm] 
    \end{array}
\end{equation}

\subsection{The Amplitudes}

Let us begin by writing the amplitudes for the two diagrams. We have
(in this case it is safe to neglect the electron mass in the propagator)
\begin{align}
  \label{eq:3}
  \mathcal{M}_t =& - \frac{g^2}{2} \overline{v}(p_2) \gamma_\nu
  (\slash{p}_1 - \slash{q}_1) \gamma_\mu u(p_1) \frac{1}{t}\
  \varepsilon^{\mu}_L(q_1)  \varepsilon^{\nu}_L(q_2) 
\nonumber\\[+2mm]
  \mathcal{M}_s =&  - \frac{g^2}{2} \overline{v}(p_2) \gamma_\alpha
  P_L u(p_1) \left[ -g^{\alpha\beta} + \frac{(p_1+p_2)^\alpha
      (p_1+p_2)^\beta}{M_W^2/c_W^2}\right]
  \frac{\Gamma_{\mu\nu\beta}(-q_1,-q_2,q_1+q_2)}{s-M_W^2/c_W^2} \ 
\varepsilon^{\mu}_L(q_1)  \varepsilon^{\nu}_L(q_2) 
\nonumber\\[+1mm]
={}& 
 \frac{g^2}{2} \overline{v}(p_2) \gamma^\alpha
  P_L u(p_1)  \frac{\Gamma_{\mu\nu\alpha}(-q_1,-q_2,q_1+q_2)}{s-M_W^2/c_W^2} \
\varepsilon^{\mu}_L(q_1)  \varepsilon^{\nu}_L(q_2) 
\end{align}
where we have used the \texttt{SM} relation $M_W=M_Z c_W$ with
$c_W=\cos\theta_W$ and, in the last step, we have used the Dirac
equation and the fact that the neutrinos (in the \texttt{SM}) have no
mass to simplify the numerator of the $Z$ propagator. 

\subsection{Unitarity and the cancellation of the bad behavior}

We can convince ourselves easily that each of the amplitudes in
Eq.~(\ref{eq:3}) do not obey the constraint from unitarity,
Eq.~(\ref{eq:14}). To see that one has to understand that at high
energy we have
\begin{equation}
  \label{eq:21}
  \overline{v}(p_2) u(p_1) \propto \sqrt{s}, \quad p_i,q_i \propto
  \sqrt{s}, \quad \varepsilon^{\mu}_L(q_i) \propto \frac{\sqrt{s}}{M_W}
\end{equation}
and therefore 
\begin{equation}
  \label{eq:22}
  \mathcal{M}_t, \mathcal{M}_s \propto s
\end{equation}
Then, unless there is a cancellation between the fastest growing terms
with $\sqrt{s}$, we will have a problem. This cancellation comes about
because gauge invariance of the theory forces the couplings to be
related. To see this let us start by the simplest formula for the
longitudinal polarization, Eq.~(\ref{eq:8}). Using this relation we
obtain
\begin{equation}
  \label{eq:23}
\mathcal{M}_{t}=\frac{1}{2M_W^2} \overline{v}(p_2)\slash{q}_2 P_L
u(p_1) + \mathcal{O}\left( 1 \right) \ .
\end{equation}
which indeed shows that the amplitude grows with $s$. If we do the
same for the s-channel diagram of the $Z$, we obtain,
\begin{align}
  \label{eq:24}
  \mathcal{M}_{s}={}&\frac{1}{4M_W^2} \overline{v}(p_2)\slash{q}_1 P_L
u(p_1) - \frac{1}{4M_W^2} \overline{v}(p_2)\slash{q}_2 P_L
u(p_1) + \mathcal{O}\left( 1 \right) \nonumber \\[+2mm]
=&  - \frac{1}{2M_W^2} \overline{v}(p_2)\slash{q}_2 P_L
u(p_1) + \mathcal{O}\left( 1 \right)
\end{align}
where we have used $q_1=-q_2+p_1+p_2$ and the Dirac equation for
massless neutrinos. We therefore obtain,
\begin{equation}
  \label{eq:25}
  \mathcal{M}= \mathcal{M}_{\nu}+ \mathcal{M}_{Z} = \mathcal{O}\left( 1 \right)
\end{equation}
in agreement with Eq.~(\ref{eq:14}). It should be stressed that the
cancellation depends on the relation between the couplings of
different particles and these were dictated by gauge invariance. 

\subsection{Cross section}

\begin{figure}[htb]
  \centering
  \begin{tabular}{cc}
    \includegraphics[width=0.4\linewidth]{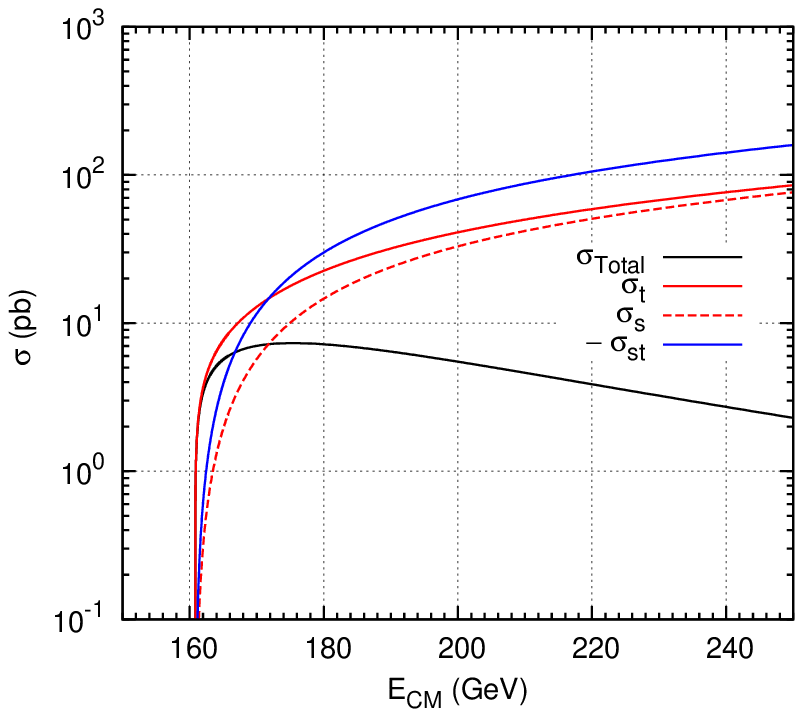}&
    \includegraphics[width=0.4\linewidth]{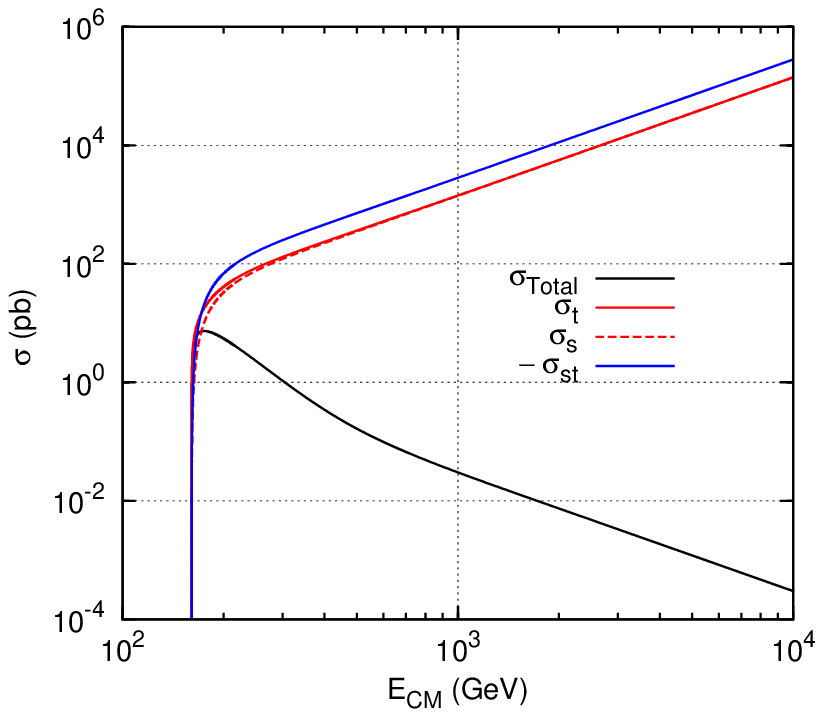}
  \end{tabular}
\vspace{-5mm}
  \caption{Cross section for the process $\nu_e + \overline{\nu}_e
    \rightarrow W^+_L  +W^-_L $. Displayed are the total cross section
  (black), as well as the s-channel (red-dashed), t-channel (red) and the
  interference (blue). As the interference is negative we plot its
  absolute value. On the left panel the range of $\sqrt{s}$ is
  smaller to show the details at threshold. On the right panel the
  behavior at high energy $\sqrt{s}\gg M_W$, is already present. Notice
the log scale on $\sqrt{s}$ on the right panel. } 
  \label{fig:xs-nunuWLWL}
\end{figure}

Let us calculate the cross section to display the effect of the
cancellation.
Writing, in an obvious notation,
\begin{equation}
  \label{eq:26}
  |\mathcal{M}|^2= |\mathcal{M}_t|^2 +|\mathcal{M}_s|^2 + 
\left(\mathcal{M}_t \mathcal{M}_s^* + \mathcal{M}_t^* \mathcal{M}_s  \right)
\end{equation}
we get (see Appendix~\ref{ap:nunuWLWL} for an input file for the
\texttt{Mathematica} package \texttt{FeynCalc}), 
\begin{align}
  \label{eq:27}
  |\mathcal{M}_t|^2 =&-\frac{g^4 \left(4 M_W^4+s
   t\right)^2 \left[M_W^4-2
   M_W^2 t+t (s+t)\right]}{4
   M_W^4 t^2 \left(s-4
   M_W^2\right)^2}\nonumber \\[+2mm]
={} & g^4 \left[ \sin^2\theta\ x^2
+  \sin^2\theta\ x -
   \cos^2\theta -\cos\theta + \mathcal{O}\left( x^{-1}\right) \right]
\end{align}
where we have defined $\displaystyle x=\frac{s}{4 M_W^2}$ and the
expansion is for $x \gg 1$.  Similarly for the s-channel $Z$ exchange
we get,
\begin{align}
  \label{eq:29}
  |\mathcal{M}_s|^2 =&-\frac{g^4 \left(2
   M_W^2+s\right)^2
   \left(M_W^4-2 M_W^2 t+t
   (s+t)\right)}{4 M_W^4
   \left(M_Z^2-s\right)^2}\nonumber \\[+2mm]
={} &
  g^4\left[\sin^2\theta\  x^2
+\frac{M_Z^2 \sin^2\theta }{2M_W^2}\ x +
\frac{3 M_Z^4 \sin^2\theta}{16M_W^4}
-\frac{3 \sin^2\theta}{4} + \mathcal{O}\left( x^{-1}\right) \right]
\end{align}
and for the interference term
\begin{align}
  \label{eq:30}
  \left(\mathcal{M}_t \mathcal{M}_s^* + \mathcal{M}_t^* \mathcal{M}_s
  \right) ={} & \frac{g^4 \left(8 M_W^{10}+4
   M_W^8 (s-4 t)+2 M_W^6 t
   (s+4 t)+5 M_W^4 s^2 t+2
   M_W^2 s t^3+s^2 t^2
   (s+t)\right)}{2 M_W^4 t
   \left(4 M_W^2-s\right)
   \left(M_Z^2-s\right)}\\[+2mm]
= {} &
g^4\left[
-2 \sin^2\theta\ x^2 
-\sin^2\theta\ x
-\frac{M_Z^2 \sin^2\theta }{2 M_W^2}\ x
+ \cos\theta +1
-\frac{M_Z^4 \sin^2\theta}{8 M_W^4}
-\frac{M_Z^2 \sin^2\theta}{4 M_W^2}
+ \mathcal{O}\left( x^{-1}\right) \right]\nonumber 
\end{align}
One can see by looking at Eqs.~(\ref{eq:27}), (\ref{eq:29}) and
(\ref{eq:30}) that the terms proportional to $x^2$ and $x$ exactly
cancel and that, at high energy the behavior is
\begin{equation}
  \label{eq:31}
  |\mathcal{M}|^2=\frac{g^4 \sin^2\theta \left(M_Z^2-2
   M_W^2\right)^2}{16 M_W^4} +  \mathcal{O}\left( x^{-1}\right)
\end{equation}
consistent with Eq.~(\ref{eq:14}). The behavior of the various terms
is shown in Fig.~\ref{fig:xs-nunuWLWL}. 
We leave as an exercise to the reader to check that introducing
Eq.~(\ref{eq:31}) in Eq. (\ref{eq:13}) one can reproduce the high
energy behavior of the right panel of Fig.~\ref{fig:xs-nunuWLWL} with
a relative error below $10^{-2}$ already for $\sqrt{s}=260$ GeV and below
$10^{-4}$ for $\sqrt{s}=500$ GeV.

\section{The scattering of longitudinal gauge bosons: $W^-_L (p_1) + W^+_L (p_2) \rightarrow W^-_L (q_1) +W^+_L (q_2) $}
\label{sec:WLWLWLWL}

\begin{figure}[hbt]
  \centering
\psfrag{h}{$h$}
\psfrag{Wm}{$W^-$}
\psfrag{Wp}{$W^+$}
\psfrag{gZ}{$\gamma,Z$}
  \includegraphics[width=\linewidth]{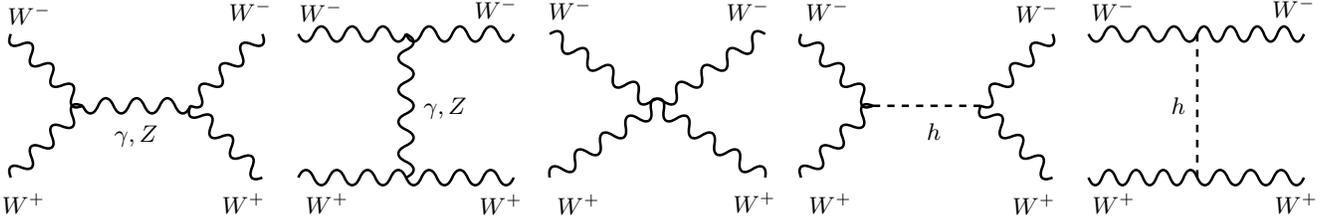}
\vspace{-2mm}
  \caption{Diagrams contributing to $W^-_L + W^+_L  \rightarrow W^-_L  +W^+_L $.}
  \label{fig:2}
\end{figure}

The next process that we will consider is the scattering of
longitudinal $W^\pm_L$.
\begin{equation}
\label{eq:1}
  W^-_L (p_1) + W^+_L (p_2) \rightarrow W^-_L (q_1) +W^+_L (q_2) 
\end{equation}
where the momenta are as indicated and the subscript $L$ means that
the gauge bosons $W^\pm$ are longitudinally polarized.
In the \texttt{SM} this process has seven tree-level diagrams shown in
Fig.~\ref{fig:2}.
with the kinematics,
\begin{equation}
  \label{eq:28}
  \left\{
    \begin{array}{l}\displaystyle
      p_1=\frac{\sqrt{s}}{2} (1, 0,0,\beta)\\[+2mm]
      \displaystyle
      p_2=\frac{\sqrt{s}}{2} (1, 0,0,-\beta)\\[+2mm]
      \displaystyle
      q_1=\frac{\sqrt{s}}{2} (1, \beta\sin\theta_{\rm
        CM},0,\beta\cos\theta_{\rm CM})\\[+2mm] 
      \displaystyle
      q_2=\frac{\sqrt{s}}{2} (1, -\beta\sin\theta_{\rm
        CM},0,-\beta\cos\theta_{\rm CM})\\[+2mm] 
    \end{array}
  \right.
\qquad
  \left\{
    \begin{array}{l}\displaystyle
      \varepsilon_L(p_1)=\frac{\sqrt{s}}{2 M_W} (\beta, 0,0,1)\\[+2mm]
      \displaystyle
      \varepsilon_L(p_2)=\frac{\sqrt{s}}{2 M_W} (\beta, 0,0,-1)\\[+2mm]
      \displaystyle
      \varepsilon_L(q_1)=\frac{\sqrt{s}}{2 M_W} (\beta, \sin\theta_{\rm
        CM},0,\cos\theta_{\rm CM})\\[+2mm] 
      \displaystyle
      \varepsilon_L(q_2)=\frac{\sqrt{s}}{2 M_W} (\beta, -\sin\theta_{\rm
        CM},0,-\cos\theta_{\rm CM})\\[+2mm] 
    \end{array}
  \right.
\end{equation}
where, as before, $\beta=\sqrt{1-4M_W^2/s}$. Notice that the invariant
relations of the type $\varepsilon_L(p_1)\cdot \varepsilon_L(p_1)=-1$
and $\varepsilon_L(p_1)\cdot p_1=0$ are verified for all cases.
We will see that this case is very interesting, because not only the
gauge structure is necessary for the amplitudes to have the correct
behavior, as in the previous case, but also the Higgs boson is fundamental.

\subsection{The Amplitudes}

\noindent
Let us denote, in an obvious notation, the amplitudes as
\begin{equation}
  \label{eq:2}
  \mathcal{M}= \mathcal{M}_{\gamma +Z}^s + \mathcal{M}_{ \gamma +Z}^t + \mathcal{M}_{4W} + \mathcal{M}_{H}^{s+t}
\end{equation}
We have then,
\begin{align}
  \label{eq:32}
  \mathcal{M}_{\gamma}^s = {} &
\frac{g^2 s_W^2}{s}\ \epsilon_L^{\alpha}(p_1) \epsilon_L^{\beta}(p_2)
\epsilon_L^{\gamma}(q_1) \epsilon_L^{\delta}(q_2)\
\Gamma_{\alpha,\beta,\mu}(p_1,p_2,-p_1-p_2) 
\Gamma_{\delta,\gamma,\nu}(-q_2,-q_1,p_1+p_2)\ g^{\mu\nu}
\nonumber\\[+2mm]
  \mathcal{M}_{Z}^s = {} &
\frac{g^2 c_W^2}{s-M_W^2/c_W^2}\ \epsilon_L^{\alpha}(p_1) \epsilon_L^{\beta}(p_2)
\epsilon_L^{\gamma}(q_1) \epsilon_L^{\delta}(q_2)\
\Gamma_{\alpha,\beta,\mu}(p_1,p_2,-p_1-p_2) 
\Gamma_{\delta,\gamma,\nu}(-q_2,-q_1,p_1+p_2) 
\nonumber\\[+1mm]
&\hskip 1mm\times
\left[g^{\mu\nu}
-\frac{(p_1+p_2)^\mu (p_1+p_2)^\nu}{M_W^2/c_W^2}\right]
\nonumber\\[+2mm]
  \mathcal{M}_{\gamma}^t = {} &
\frac{g^2 s_W^2}{t}\ \epsilon_L^{\alpha}(p_1) \epsilon_L^{\beta}(p_2)
\epsilon_L^{\gamma}(q_1) \epsilon_L^{\delta}(q_2)\
\Gamma_{\alpha,\gamma,\mu}(p_1,-q_1,q_1-p_1) 
\Gamma_{\delta,\beta,\nu}(-q_2,p_2,q_2-p_2)\ g^{\mu\nu}
\nonumber\\[+2mm]
  \mathcal{M}_{Z}^t = {} &
\frac{g^2 c_W^2}{t-M_W^2/c_W^2}\ \epsilon_L^{\alpha}(p_1) \epsilon_L^{\beta}(p_2)
\epsilon_L^{\gamma}(q_1) \epsilon_L^{\delta}(q_2)\
\Gamma_{\alpha,\gamma,\mu}(p_1,-q_1,q_1-p_1) 
\Gamma_{\delta,\beta,\nu}(-q_2,p_2,q_2-p_2)
\nonumber\\[+2mm]
 {} &\hskip 1mm\times
\left[g^{\mu\nu}
-\frac{(p_1-q_1)^\mu (p_1-q_1)^\nu}{M_W^2/c_W^2}\right]
\\[+2mm]
  \mathcal{M}_{4W} = {} &
g^2 \ \epsilon_L^{\alpha}(p_1) \epsilon_L^{\beta}(p_2)
\epsilon_L^{\gamma}(q_1) \epsilon_L^{\delta}(q_2)\
\left[ 2 g_{\alpha\delta}\, g_{\beta\gamma}- g_{\alpha\beta}\,
  g_{\delta\gamma} - g_{\alpha\gamma}\, g_{\delta\beta}\right]
\nonumber\\[+2mm]
  \mathcal{M}_{H}^s = {} &
-\frac{g^2 M_W^2}{s-M_H^2}\ 
\epsilon_L^{\alpha}(p_1) \epsilon_L^{\beta}(p_2)
\epsilon_L^{\gamma}(q_1) \epsilon_L^{\delta}(q_2)\
g_{\alpha\beta}\, g_{\gamma\delta}
\nonumber\\[+2mm]
  \mathcal{M}_{H}^t = {} &
-\frac{g^2 M_W^2}{t-M_H^2}\ 
\epsilon_L^{\alpha}(p_1) \epsilon_L^{\beta}(p_2)
\epsilon_L^{\gamma}(q_1) \epsilon_L^{\delta}(q_2)\
g_{\alpha\gamma}\, g_{\beta\delta}
\nonumber
\end{align}
where we have introduced the shorthand notation
$s_W^2=\sin^2\theta_W$, $c_W^2=\cos^2\theta_W$, and made use of the
\texttt{SM}  relations $M_W=c_W M_Z$ and $e = g s_W$. If we insert the
high energy behavior of $\epsilon_L$, Eq.~(\ref{eq:8}), we see that
the amplitudes can grow potentially as $s^2$ or even $s^3$ in the case
of the diagrams where there is a $Z$ exchange. Therefore it is clear that
the approximation of Eq.~(\ref{eq:8}) is not enough. So, we start by
calculating the amplitudes with no approximation, making use of
Eq.~(\ref{eq:12}) and of the \texttt{CM} kinematics of
Fig.~\ref{fig:CM}. We get,
\begin{align}
  \label{eq:33}
  \mathcal{M}_{\gamma}^s =&
\frac{ g^2s_W^2}{4
   M_W^4 s} \left(2
   M_W^2+s\right)^2 \left(4
   M_W^2-s-2 t\right)
\nonumber\\[+2mm]
  \mathcal{M}_{Z}^s =&
\frac{g^2\, c_W^2 }{4 M_W^4
   \left(s-M_W^2/c_W^2\right)} \left(2
   M_W^2+s\right)^2 \left(4
   M_W^2-s-2 t\right)
\nonumber\\[+2mm]
  \mathcal{M}_{\gamma}^t =&
\frac{g^2 s_W^2 }{4 M_W^4 t \left(s-4  M_W^2\right)^2} 
\left[\vb{12}256 M_W^{10}-64 M_W^8 (4 s+t)+16 M_W^6 s (5 s+14 t)-4
   M_W^4 s \left(2 s^2+21 s t+20   t^2\right)\right.
\nonumber\\[+1mm]
&\hskip 30mm\left.\vb{12}
+8 M_W^2 s^2 t (s+3 t)-s^2 t^2 (2 s+t)
\right]
\nonumber\\[+2mm]
  \mathcal{M}_{Z}^t =&
\frac{g^2 c_W^2 }{4  M_W^4 \left(s-4  M_W^2\right)^2 \left(t-M_W^2/c_W^2\right)}
   \left[\vb{12} 256  M_W^{10}-64 M_W^8 (4 s+t)+16 M_W^6 s (5 s+14 t)
     -4 M_W^4 s \left(2 s^2+21 s t+20 t^2\right)\right.
\nonumber\\[+1mm]
&\hskip 50mm\left.\vb{12}
+8 M_W^2 s^2 t (s+3 t)-s^2 t^2 (2 s+t)\right]
\\[+2mm]
  \mathcal{M}_{4W} =&
\frac{g^2  s }{4 M_W^4\left(s-4 M_W^2\right)^2}
\left[\vb{12}-64 M_W^6+48
   M_W^4 (s+t)-4 M_W^2 s (3
   s+7 t)+s \left(s^2+4 s
   t+t^2\right)\right]
\nonumber\\[+2mm]
  \mathcal{M}_{H}^s =& 
-g^2\frac{\left(s-2 M_W^2\right)^2}{4 M_W^2 \left(s-M_H^2\right)}
\nonumber\\[+2mm]
  \mathcal{M}_{H}^t =&
 - g^2\, \frac{\left(-8 M_W^4+2
   M_W^2 s+s t\right)^2}{4
   \left(t-M_H^2\right)
   \left(M_W s-4
   M_W^3\right)^2}
\nonumber
\end{align}

\subsection{Unitarity and the cancellation of the bad behavior}

By inspection, we see that for $\sqrt{s}\gg M_W$ the first five
amplitudes grow as $s^2$ and the last two (from the Higgs exchange) as $s$. So
if we define, as before, the dimensionless variable $x=s/(4M_W^2)$,
for $x\gg 1$ we should be able to write all amplitudes in the form
\begin{equation}
  \label{eq:16}
  \mathcal{M}_i= A_i x^2 + B_i x + C_i + \mathcal{O}(1/x)\ .
\end{equation}
The results for these coefficients are summarized in Table~\ref{tab:1}.
\begin{table}[htb]
  \centering
  \begin{tabular}{|c|c|c|c|}\hline
\vb{10}    &$A_i$& $B_i$ & $C_i$\\[+1mm]\hline 
\vb{10} $\mathcal{M}_{\gamma}^s$ &$-g^2 4 s_W^2\cos\theta$ &$0$ & $g^2 3 s_W^2\cos\theta$\\[+2mm]
 $\mathcal{M}_{Z}^s$ &$-g^2 4 c_W^2\cos\theta$ &$-g^2\cos\theta$ &
$g^2 \left[3 \cos\theta\ c_W^2-\displaystyle\frac{\cos\theta}{4 c_W^2}\right]$
\\[+2mm]
\vb{12} $\mathcal{M}_{\gamma}^t$ &
$g^2\, s_W^2 \left(\vb{10}-\cos^2\theta -2
   \cos\theta +3 \right)$
&$g^2\, 8 s_W^2  \cos\theta $ 
&$\displaystyle\frac{g^2 s_W^2}{\cos\theta-1} \left(\vb{10} - 2 \cos^2\theta
   -\cos\theta - 1\right)$
\\[+2mm]
\vb{12} 
$\mathcal{M}_{Z}^t$ 
& $g^2\, c_W^2\left(\vb{10} -\cos^2\theta -2
   \cos\theta +3 \right)$
&$g^2\left(8 \cos\theta
   c_W^2- \displaystyle\frac{\cos\theta}{2}-\displaystyle\frac{3}{2}\right)$
 &$\displaystyle\frac{g^2}{\cos\theta-1}\left[\vb{10}-2 \cos^2\theta
   c_W^2-\frac{\cos^2\theta}{2}
   -\cos\theta
   c_W^2-\frac{\cos\theta}{4
   c_W^2}\right.$
\\[+1mm]
\vb{12} 
& &
 &$\left.\vb{12} +3 \cos\theta-c_W^2-\frac{3}{4
   c_W^2}+\displaystyle\frac{3}{2}\right]$
\\[+2mm]
\vb{12} $\mathcal{M}_{4W}$ 
& $g^2\left(\vb{10}\cos^2\theta+6 \cos\theta-3\right)$
&$g^2\left(2-6 \cos\theta\right)$ 
&$0$\\[+2mm]\hline
\vb{14}$\displaystyle\sum_{\gamma Z}$
&$0$&$g^2 \displaystyle
\frac{1+\cos\theta}{2}$&$g^2 3\cos\theta + \cdots$\\[+2mm]\hline
\vb{14}  $\mathcal{M}_{H}^s$ &$0$ &$-g^2$ 
&$g^2\left(1-\displaystyle\frac{M_H^2}{4 M_W^2}\right)$\\[+3mm]
\vb{14}  $\mathcal{M}_{H}^t$ &$0$ 
&$g^2\left(\displaystyle\frac{1}{2}-\displaystyle\frac{\cos\theta}{2}\right)$ 
&$-g^2 \left[
\displaystyle\frac{1+\cos\theta}{2}+\displaystyle\frac{M_H^2}{4 M_W^2}
\right]  $
\\[+3mm]\hline
\vb{14}$\displaystyle\sum_{\gamma Z H}$
&$0$&$0$&$\not=0$\\[+2mm]\hline
  \end{tabular}
  \caption{Coefficients $A_i$, $B_i$ and $C_i$ (see text for an explanation).}
  \label{tab:1}
\end{table}
We see that the terms proportional to $x^2$ cancel among the first
five diagrams involving only the gauge bosons, but that the term
proportional do $x$ remains after we sum over the gauge part. So, 
if we  consider only a gauge theory of intermediate gauge bosons, we
are in trouble. This trouble can be traced back to the fact that with
mass the gauge invariance is lost, and the theory is inconsistent if
the diagrams involving the Higgs boson field are not taken in
account. In conclusion, again the Higgs boson is crucial to make the
\texttt{SM} consistent. 

\subsection{Cross section}

\begin{figure}[htb]
  \centering
  \includegraphics[width=0.45\linewidth]{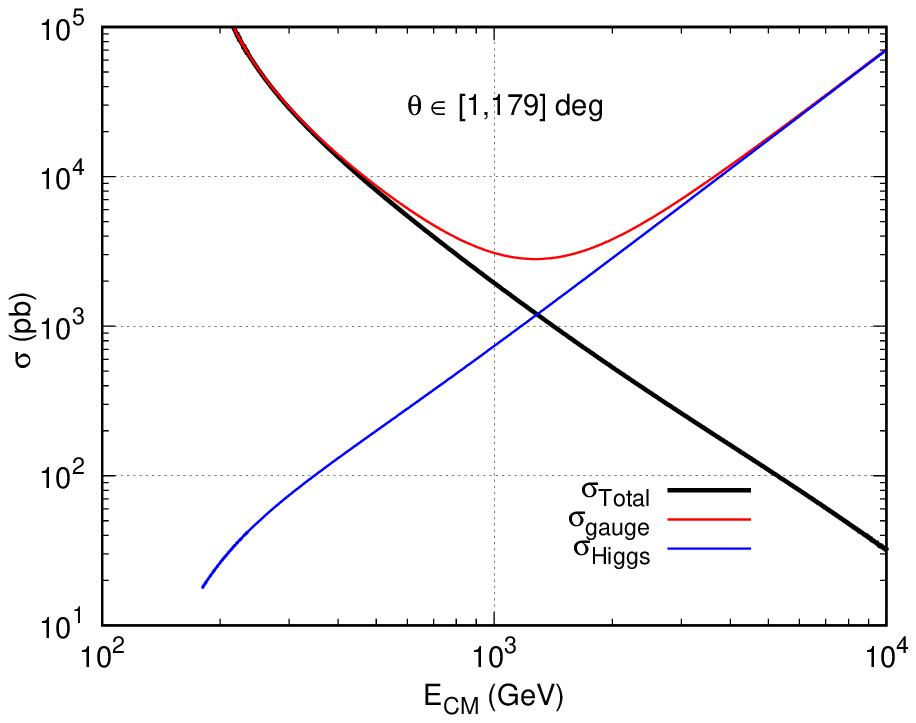}
    \includegraphics[width=0.45\linewidth]{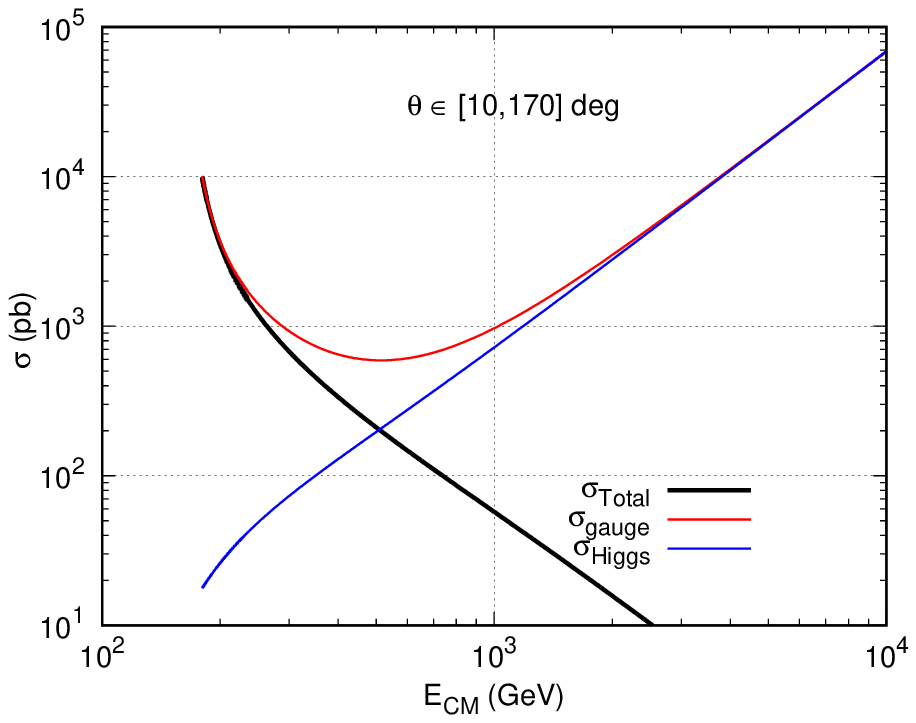}
  \caption{Cross section for $W^-_L + W^+_L  \rightarrow W^-_L  +W^+_L
    $. Shown are the contribution of the gauge diagrams (red), the
    contribution from the Higgs (blue) and the total cross section
    (black). The sum of the amplitudes from the gauge part have the
    opposite sign from those from the Higgs (not visible in the Fig,
    because we are plotting cross sections) forcing the cross section
    to decrease. On the left panel we consider $\theta_{\rm
  min}=1^\circ$ while in the right panel we take $\theta_{\rm
  min}=10^\circ$.}
  \label{fig:WLWLWLWL}
\end{figure}

As, in this case, the amplitudes are pure c-numbers with no spinor part,
the cross section is simply obtained by summing all the amplitudes and
taking the absolute value of the result to obtain
$\mathcal{M}|^2$. The cancellation of the bad high energy behavior is 
shown in Fig.~\ref{fig:WLWLWLWL}. As it is well known, there is a pole
in the t-channel of the photon. To avoid that we cut in the scattering
angle. Shown are the plots for two cases, $\theta_{\rm
  min}=1^\circ,10^\circ$. The cancellation is clear in both cases.

\section{The scattering  $e^- +e^+ \rightarrow W^-_L  +W^+_L $}
\label{sec:eEWLWL}

\begin{figure}[htb]
  \centering
\psfrag{h}{$h$}
\psfrag{n}{$\nu_e$}
\psfrag{Wp}{$W^-$}
\psfrag{Wm}{$W^+$}
\psfrag{e}{$e^-$}
\psfrag{p}{$e^+$}
\psfrag{g}{$\gamma$}
\psfrag{z}{$Z$}
  \includegraphics{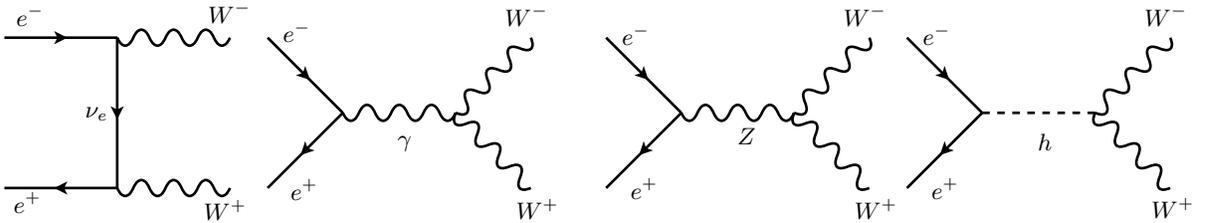}
  \caption{Diagrams contributing to $e^- + e^+ \rightarrow W^-_L  +W^+_L$.}
  \label{fig:3}
\end{figure}

As a last example of the importance of the Higgs boson for the
consistency of the \texttt{SM} we consider the process,
\begin{equation}
  \label{eq:35}
  e^-(p_1) +e^+(p_2) \rightarrow W^-_L(q_1) +W^+_L(q_2)
\end{equation}
Compared to other two, this process has
the advantage of not being an academic problem, it has in fact already
been tested at the LEP experiments. In the \texttt{SM} this process
has four tree-level diagrams shown in Fig.~\ref{fig:3}.
to which correspond the the following kinematics,
\begin{equation}
  \label{eq:28}
  \left\{
    \begin{array}{l}\displaystyle
      p_1=\frac{\sqrt{s}}{2} (1, 0,0,\beta_e)\\[+2mm]
      \displaystyle
      p_2=\frac{\sqrt{s}}{2} (1, 0,0,-\beta_e)\\[+2mm]
      \displaystyle
      q_1=\frac{\sqrt{s}}{2} (1, \beta\sin\theta_{\rm
        CM},0,\beta\cos\theta_{\rm CM})\\[+2mm] 
      \displaystyle
      q_2=\frac{\sqrt{s}}{2} (1, -\beta\sin\theta_{\rm
        CM},0,-\beta\cos\theta_{\rm CM})\\[+2mm] 
    \end{array}
  \right.
\qquad
  \left\{
    \begin{array}{l}\displaystyle
      \varepsilon_L(q_1)=\frac{\sqrt{s}}{2 M_W} (\beta, \sin\theta_{\rm
        CM},0,\cos\theta_{\rm CM})\\[+2mm] 
      \displaystyle
      \varepsilon_L(q_2)=\frac{\sqrt{s}}{2 M_W} (\beta, -\sin\theta_{\rm
        CM},0,-\cos\theta_{\rm CM})\\[+2mm] 
    \end{array}
  \right.
\end{equation}
where, as before, $\beta=\sqrt{1-4M_W^2/s}$ and, (we keep $m_e\not= 0$),
 $\beta_e=\sqrt{1-4m_e^2/s}$.

\subsection{The Amplitudes}
We start by writing the amplitudes for the four diagrams. As the Higgs
coupling to the electrons is proportional to the electron mass, we
expect, from the other examples, that for sufficiently high energy
there will be a piece of the first three diagrams proportional to
$m_e$ that will increase with energy. Therefore, we are not making any
approximation with regard to $m_e$, and we should keep it in the
calculations. We get
\begin{align}
  \label{eq:34}
  \mathcal{M}_{\nu}^t=&
-\frac{g^2}{2 t} \overline{v}(p_2) \gamma_\nu P_L \left( \slash{p}_1
  -\slash{q}_1 \right)\gamma_\mu P_L u(p_1)\ \epsilon_L^{\mu}(q_1)
  \epsilon_L^{\nu}(q_2) 
\nonumber \\[+2mm]
  \mathcal{M}_{\gamma}^s=&
- \frac{g^2 s_W^2}{s} \overline{v}(p_2) \gamma^\alpha  u(p_1)\ 
  \Gamma_{\nu\mu\alpha}(-q_2,-q_1,q_1+q_2)\ \epsilon_L^{\mu}(q_1)
  \epsilon_L^{\nu}(q_2) 
\nonumber \\[+2mm]
  \mathcal{M}_{Z}^s=&
- \frac{g^2 }{s-M_W^2/c_W^2}\left[-g_{\alpha\beta} +
  \frac{Q^\alpha Q^\beta}{M_W^2/c_W^2} \right]
  \overline{v}(p_2) \gamma^\beta \left( g_L P_L +g_R P_R \right) u(p_1)\ 
  \Gamma_{\nu\mu\alpha}(-q_2,-q_1,q_1+q_2)\ \epsilon_L^{\mu}(q_1)
  \epsilon_L^{\nu}(q_2) 
\nonumber \\[+2mm]
  \mathcal{M}_{H}^s=&
\frac{g^2 m_e}{2}\frac{1}{s-M_H^2}\ \overline{v}(p_2)  u(p_1)\ g_{\mu\nu}\
 \epsilon_L^{\mu}(q_1) \epsilon_L^{\nu}(q_2) 
\end{align}
where we have defined $Q=p_1+p_2$, and made use again of the
\texttt{SM} relations like $e=g s_W$. If we neglect for the moment
$m_e$ this process is very much like the neutrino scattering studied
in section~\ref{sec:nunuWLWL}. Using Eq.~(\ref{eq:21}) one can
convince ourselves that the first three amplitudes are proportional to
$s$ for $\sqrt{s}\gg m_e, M_W$. However we should be careful, because
the Higgs exchange diagram is proportional to $m_e\sqrt{s}$ and that
these terms, although sub-leading, will be important for sufficiently
high energy.

\subsection{Cancellation of bad high energy behavior}

Let us now see how the cancellation of the bad high energy behavior is
achieved in this case. One can check that it is enough to use the
approximate formula of Eq.~(\ref{eq:8}) to display the cancellation of
the diagrams. With the help of \texttt{FeynCalc} we obtain
\begin{align}
  \mathcal{M}_\nu=&\frac{g^2}{2 M_W^2}\ \left[\vb{12} - \overline{v}(p_2)
    \slash{q}_1 P_L u(p_1) - m_e  \overline{v}(p_2) P_L  u(p_1)
    \right] + \mathcal{O}(1/x) \nonumber\\[+2mm]
\mathcal{M}_\gamma=&\frac{g^2 s_W^2}{M_W^2}\ \overline{v}(p_2)
 \slash{q}_1 \left( P_L + P_R\right) u(p_1)+ \mathcal{O}(1/x)
 \nonumber\\[+2mm] 
\mathcal{M}_Z=& - \frac{g^2 s_W^2}{M_W^2}\ \overline{v}(p_2)
 \slash{q}_1 \left( P_L + P_R\right) u(p_1) + \frac{g^2 }{2M_W^2}\
\overline{v}(p_2) \slash{q}_1 P_L u(p_1) \nonumber\\[+2mm]
{}& +\frac{g^2 }{4M_W^2}\ m_e\left[\vb{12} \overline{v}(p_2) P_L  u(p_1) 
- \overline{v}(p_2) P_R  u(p_1) \right] + \mathcal{O}(1/x)
\nonumber\\[+2mm]
\mathcal{M}_H=&
\frac{g^2 }{4M_W^2}\ m_e \left[\vb{12}\overline{v}(p_2) P_L  u(p_1) 
+ \overline{v}(p_2) P_R  u(p_1) \right] + \mathcal{O}(1/x)
\end{align}
and therefore
\begin{equation}
  \label{eq:42}
  \mathcal{M}_\nu+ \mathcal{M}_\gamma + \mathcal{M}_Z
= - \frac{g^2 }{4M_W^2}\ m_e \left[\vb{12}\overline{v}(p_2) P_L  u(p_1) 
+ \overline{v}(p_2) P_R  u(p_1) \right] + \mathcal{O}(1/x)
= - \mathcal{M}_H
\end{equation}
Hence the Higgs boson exchange is needed to cancel the contribution of
the amplitude that grows like $\displaystyle \frac{m_e
  \sqrt{s}}{M_W^2}$. 

\subsection{Cross section}

\begin{figure}[htb]
   \centering
   \begin{tabular}{cc}
     \includegraphics[width=0.45\linewidth]{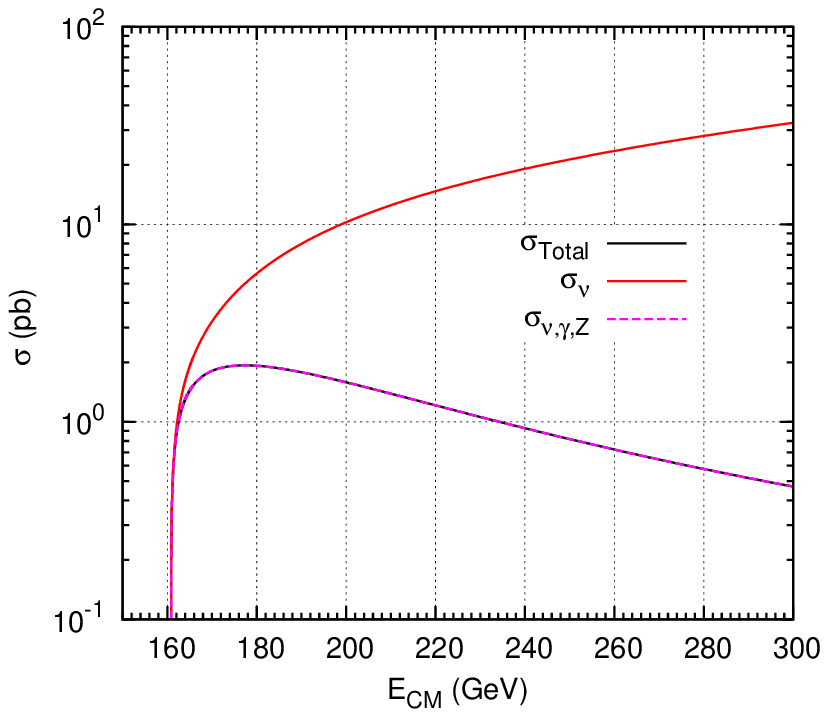}
     \includegraphics[bb=0 60 567 624,width=0.45\linewidth]{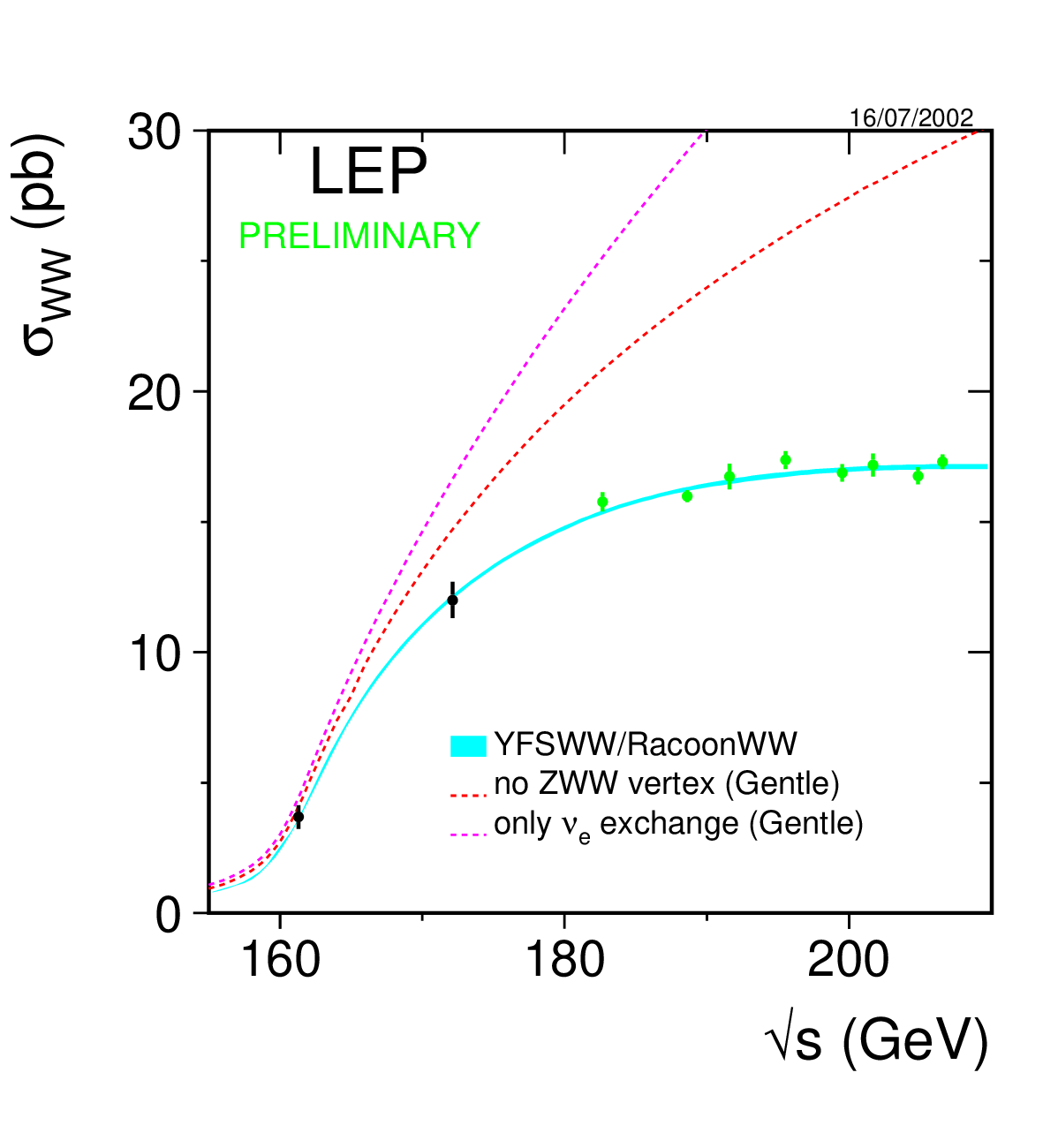}
   \end{tabular}
   \caption{On the left panel it is shown the total cross section for
     $e^- + e^+ \rightarrow W^-_L +W^+_L $ (black). Also shown are the
   $\nu$ exchange cross section (red) and the sum of the $\nu$
   exchange with the s-channel exchange of gauge bosons (magenta).
On the right panel it is shown the LEP result for $e^- +
   e^+ \rightarrow W^- +W^+ $. }
   \label{fig:wwxs}
 \end{figure}

\begin{figure}[htb]
   \centering
   \includegraphics[width=0.45\linewidth]{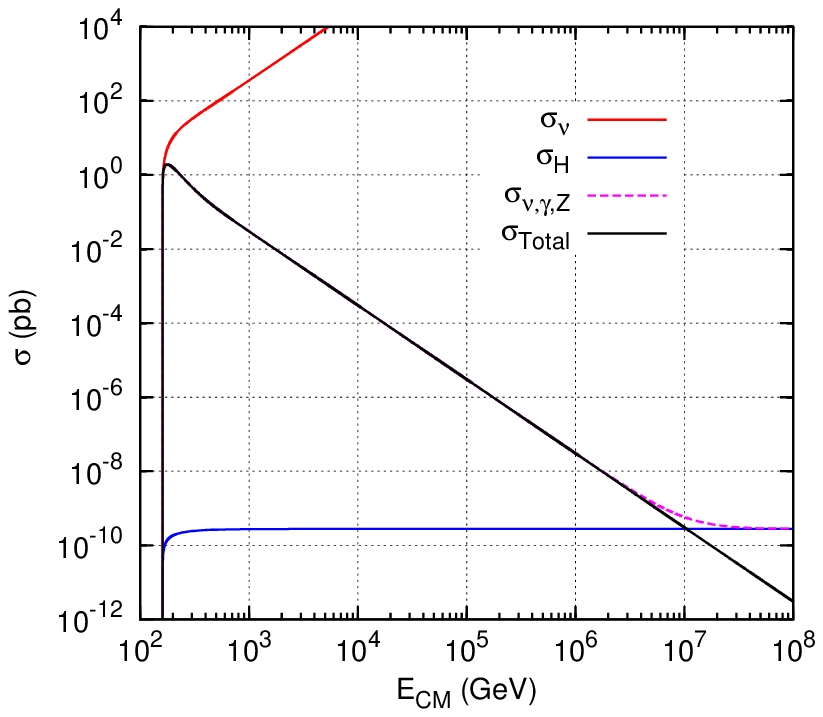}
   \caption{Various contributions to the
     cross section $e^- + e^+ \rightarrow W^-_L +W^+_L
     $ at very high energy. See text for details.}
   \label{fig:wwxs-HighEnergy}
 \end{figure}

To evaluate the cross section it is simpler to write the total
amplitude as 
\begin{equation}
  \label{eq:36}
  \mathcal{M}= g^2 \overline{v}(p_2)\Gamma_{\mu\nu} u(p_1)\
  \epsilon_L^{\mu}(q_1)   \epsilon_L^{\nu}(q_2),\qquad 
\Gamma_{\mu\nu}=\sum_i \Gamma^i_{\mu\nu}
\end{equation}
where the $\Gamma^i_{\mu\nu}$ are
\begin{align}
  \label{eq:37}
  \Gamma_{\mu\nu}^{\nu t} =&
-\frac{1}{2 t}  \gamma_\nu P_L \left( \slash{p}_1
  -\slash{q}_1 \right)\gamma_\mu P_L u(p_1)
  \nonumber\\[+2mm]
  \Gamma_{\mu\nu}^{\gamma s} =&
- \frac{ s_W^2}{s}  \gamma^\alpha  \Gamma_{\nu\mu\alpha}(-q_2,-q_1,q_1+q_2)
  \nonumber\\[+2mm]
  \Gamma_{\mu\nu}^{Z s} =&
- \frac{1 }{s-M_W^2/c_W^2}\left[-g_{\alpha\beta} +
  \frac{Q^\alpha Q^\beta}{M_W^2/c_W^2} \right]
   \gamma^\beta \left( g_L P_L +g_R P_R \right)  
  \Gamma_{\nu\mu\alpha}(-q_2,-q_1,q_1+q_2)
  \nonumber\\[+2mm]
  \Gamma_{\mu\nu}^{H s} =&
  \frac{m_e}{2}\frac{1}{s-M_H^2} g_{\mu\nu}
\end{align}
With this notation the differential cross section is given by
Eq.~(\ref{eq:13}) where
\begin{equation}
  \label{eq:38}
  |\mathcal{M}|^2 = \frac{1}{4} \text{Tr} \left[ (\slash {p}_2 -m_e)
    \Gamma (\slash{p}_1+m_e ) \overline{\Gamma} \right]
\end{equation}
where $\overline{\Gamma}=\gamma^0 \Gamma^{\dagger} \gamma^0$. With the
help of \texttt{FeynCalc} (see Appendix~\ref{ap:eEWLWL}), we evaluate
 three cases, namely neutrino cross section, the contribution from the
 sum of the neutrino and gauge bosons, and finally the total cross
 section including the Higgs boson diagram. We get,
 \begin{align}
   \label{eq:39}
   |\mathcal{M}_\nu|^2 =&
   -\frac{g^4}{16 M_W^4 t^2 \left(s-4 M_W^2\right)^2} 
  \left[\vb{12} m_e^8 s^2 -2  m_e^6 s \left(4 M_W^4+M_W^2 s+2 s t\right)
   +m_e^4 \left(16 M_W^8+16 M_W^6  s+M_W^4 s (s+24 t) \right.\right.
\nonumber\\[+1mm]
&\hskip 30mm \left.\left.\vb{12}
   +2 M_W^2 s^2 t+s^2 t (s+6 t)\right)
 -2 m_e^2 \left(16 M_W^{10}+4 M_W^8 s+8 M_W^6 s t+4 M_W^4 s t(s+3 t)
\right.\right.
\nonumber\\[+1mm]
&\hskip 30mm \left.\left.\vb{12}
-M_W^2 s^2 t^2+s^2 t^2
   (s+2 t)\right)+\left(4 M_W^4+s
   t\right)^2 \left(M_W^4-2
   M_W^2 t+t
   (s+t)\right)\right]
   \nonumber\\[+2mm]
   |\mathcal{M}_{\gamma}|^2 =&
-\frac{g^4 s_W^4}{2 M_W^4 s^2} \left(2
   M_W^2+s\right)^2
   \left[\vb{12}m_e^4+m_e^2
   \left(2 M_W^2-s-2
   t\right)+M_W^4-2 M_W^2
   t+t (s+t)\right]
   \nonumber\\[+2mm]
   |\mathcal{M}_{Z}|^2 =&
\frac{g^4 c_W^4  }{32 M_W^4
   \left(M_W^2-c_W^2
   s\right)^2} \left( 2 M_W^2+s\right)^2 
\left[-2 m_e^4 \left(8 s_W^4-4  s_W^2+1\right)
  +m_e^2 \left(4 \left(-8 M_W^2  s_W^4+4 M_W^2 s_W^2
   +8 s_W^4 t
\right.\right.\right.
 \nonumber\\[+2mm]
&\hskip 30mm
\left.\left.\left.
-4 s_W^2 t+t\right)
   +s \left(1-4 s_W^2\right)^2\right)-2 \left(8 s_W^4-4 s_W^2+1\right)
   \left(M_W^4-2 M_W^2 t+t (s+t)\right)\right]
   \nonumber\\[+2mm]
   \mathcal{M}_{\nu}\mathcal{M}_{\gamma}^* =&   
   \mathcal{M}_{\nu}^*\mathcal{M}_{\gamma}
  \nonumber\\[+2mm]
=&\frac{g^4 s_W^2  }{8 M_W^4 s t
   \left(4 M_W^2-s\right)}\left[\vb{12} 
 m_e^6  s \left(2 M_W^2+s\right)+m_e^4 \left(2 M_W^2+s\right) 
 \left(4  M_W^4-4 M_W^2 s-3 s  t\right)
 +m_e^2 s \left(6  M_W^6
\right.\right.
 \nonumber\\[+2mm]
&\hskip 30mm
\left.\left.
+M_W^4 (3 s+4 t)+2 M_W^2 t (2 s+3 t)+s t (s+3
   t)\right)-8 M_W^{10}-4 M_W^8 (s-4 t)
\right.
 \nonumber\\[+2mm]
&\hskip 30mm
\left.\vb{12}
-2 M_W^6 t (s+4 t)
   -5 M_W^4 s^2 t-2 M_W^2 s t^3-s^2 t^2 (s+t)\right]
   \nonumber\\[+2mm]
   \mathcal{M}_{\nu}\mathcal{M}_{Z}^* =&   
   \mathcal{M}_{\nu}^*\mathcal{M}_{Z} \nonumber\\[+2mm]
=&
-\frac{ g^4  c_W^2 \left(2
   M_W^2+s\right)}{16 M_W^4
   t \left(4 M_W^2-s\right)
   \left(M_W^2-c_W^2  s\right)} 
   \left[\vb{12}m_e^6 \left(s-2 s s_W^2\right)+m_e^4
   \left(-8 M_W^4 s_W^2+M_W^2 s \left(8 s_W^2-3\right)
   +3 s \left(2  s_W^2-1\right) t\right)
\right.
 \nonumber\\[+2mm]
&\hskip 30mm
\left.
   +m_e^2 \left(4 M_W^6+M_W^4 \left(-6 s s_W^2+2 s+4
   t\right)+M_W^2 s \left(1-4  s_W^2\right) t-s \left(2
   s_W^2-1\right) t (s+3  t)\right)
\right.
 \nonumber\\[+2mm]
&\hskip 30mm
\left.\vb{12}
+\left(2 s_W^2-1\right) \left(4  M_W^8-8 M_W^6 t+M_W^4 t (5 s+4 t)-2
   M_W^2 s t^2+s t^2 (s+t)\right)\right]
   \nonumber\\[+2mm]
   \mathcal{M}_{\gamma}\mathcal{M}_{Z}^* =&   
   \mathcal{M}_{\gamma}^*\mathcal{M}_{Z} 
   \nonumber\\[+2mm]
=&
-\frac{ g^4 c_W^2 s_W^2
   \left(4 s_W^2-1\right) \left(2
   M_W^2+s\right)^2
   \left[\vb{12}m_e^4+m_e^2
   \left(2 M_W^2-s-2
   t\right)+M_W^4-2 M_W^2
   t+t (s+t)\right]}{8 M_W^4 s
   \left(M_W^2-c_W^2
   s\right)}
   \nonumber\\[+2mm]
   |\mathcal{M}_{H}|^2 =&
-\frac{g^4 m_e^2 \left(4
   m_e^2-s\right) \left(s-2
   M_W^2\right)^2}{32 M_W^4
   \left(M_H^2-s\right)^2}
   \nonumber\\[+2mm]
\mathcal{M}_{\nu}\mathcal{M}_{H}^* =&   
   \mathcal{M}_{\nu}^*\mathcal{M}_{H} 
   \nonumber\\[+2mm]
=&
-\frac{g^4 m_e^2 \left(2
   M_W^2-s\right)
   \left[\vb{12}m_e^4 s+m_e^2
   \left(4 M_W^4-3 M_W^2
   s-2 s t\right)-4 M_W^6+2
   M_W^4 (s+2 t)-3 M_W^2 s
   t+s t (s+t)\right)}{16 M_W^4 t
   \left(M_H^2-s\right) \left(s-4
   M_W^2\right]}
   \nonumber\\[+2mm]
\mathcal{M}_{\gamma}\mathcal{M}_{H}^* =&   
   \mathcal{M}_{\gamma}^*\mathcal{M}_{H} =
\frac{g^4 m_e^2 s_W^2
   \left(2 M_W^2-s\right) \left(2
   M_W^2+s\right) \left(2
   m_e^2+2 M_W^2-s-2
   t\right)}{8 M_W^4 s
   \left(s-M_H^2\right)}
   \nonumber\\[+2mm]
\mathcal{M}_{Z}\mathcal{M}_{H}^* =&   
   \mathcal{M}_{Z}^*\mathcal{M}_{H} =
\frac{g^4  c_W^2 m_e^2
   \left(4 s_W^2-1\right) \left(2
   M_W^2-s\right) \left(2
   M_W^2+s\right) \left(2
   m_e^2+2 M_W^2-s-2
   t\right)}{32 M_W^4
   \left(s-M_H^2\right)
   \left(M_W^2-c_W^2
   s\right)}
    \end{align}
and the total $|\mathcal{M}|^2$ is given by
\begin{align}
  \label{eq:40}
  |\mathcal{M}|^2 =&|\mathcal{M}_\nu|^2 + |\mathcal{M}_\gamma|^2 +
|\mathcal{M}_Z|^2 + |\mathcal{M}_H|^2 +
\left(\mathcal{M}_{\nu}\mathcal{M}_{\gamma}^* +
  \mathcal{M}_{\nu}^*\mathcal{M}_{\gamma}\right) + 
\left(\mathcal{M}_{\nu}\mathcal{M}_{Z}^* +
  \mathcal{M}_{\nu}^*\mathcal{M}_{Z}\right)
+\left(\mathcal{M}_{\gamma}\mathcal{M}_{Z}^* +
  \mathcal{M}_{\gamma}^*\mathcal{M}_{Z}\right)
\nonumber\\[+2mm]
&
+\left(\mathcal{M}_{\nu}\mathcal{M}_{H}^* +
  \mathcal{M}_{\nu}^*\mathcal{M}_{H}\right) +
\left(\mathcal{M}_{\gamma}\mathcal{M}_{H}^* +
  \mathcal{M}_{\gamma}^*\mathcal{M}_{H}\right)+ 
\left(\mathcal{M}_{Z}\mathcal{M}_{H}^* +
  \mathcal{M}_{Z}^*\mathcal{M}_{H}\right) 
\end{align}
All these complicated expressions were obtained with the package
\texttt{FeynCalc} for \texttt{Mathematica} (see
Appendix~\ref{ap:eEWLWL} for details). We manipulate the
expressions using the functions \texttt{TeXForm} for \texttt{LaTeX}
  output and \texttt{FortranForm} for the \texttt{Fortran} output. In
  this way we minimize the errors of handling complicated
  expressions. The results for the cross section are shown on
  Fig.~\ref{fig:wwxs}.
On the left panel it is shown the total cross section for $e^- + e^+
\rightarrow W^-_L +W^+_L $ (black). Also shown are the $\nu$ exchange
cross section (red) and the sum of the $\nu$ exchange with the
s-channel exchange of gauge bosons (magenta). We see that it does not
differ, for this energy range, from the total cross section. On the
right panel it is shown the LEP result for $e^- + e^+ \rightarrow W^-
+W^+ $. Notice that on the right panel all the polarizations of the W
are considered. As, for these energies, the dominant contribution is
from the transverse polarized $W$'s, the numbers can not be directly
compared. However, the cancellation of the bad behavior is clear.
At these energies the contribution of the diagrams involving the Higgs
boson is very small, due to the smallness of the cross section the
electron mass and can be completed neglected at LEP energies. One can
estimate that the importance of the Higgs boson diagrams will show up
when 
\begin{equation}
  \label{eq:41}
  \frac{m_e \sqrt{s}}{M_W^2}  \simeq 1, \rightarrow \sqrt{s} \simeq 10^7\text{GeV}
\end{equation}
an energy completely outside the reach of man-made
accelerators. However, from the consistency point of view, the
cancellation of the bad behavior at those energies should be
there. This is shown in Fig.~\ref{fig:wwxs-HighEnergy}. Here we show
the contribution of the various pieces to the cross section of the
process $e^- + e^+ \rightarrow W^-_L +W^+_L $. The total cross section
is shown in black, the neutrino exchange in red and the sum of the
neutrino with the gauge bosons in magenta. In blue is the contribution
from the Higgs boson exchange. We see that for high enough energy,
$\sqrt{s}\simeq M_W^2/m_e \simeq 10^{7}$ GeV, the Higgs boson
contribution is crucial to avoid the cross section to be constant. If
the electron mass was not so small, the effect would be seen much
earlier as in the case of $W^-_L + W^+_L \rightarrow W^-_L +W^+_L $.

\subsection{Other Polarizations of the $W$'s}

\begin{figure}[htb]
  \centering
  \begin{tabular}{cc}
\includegraphics[width=0.45\linewidth]{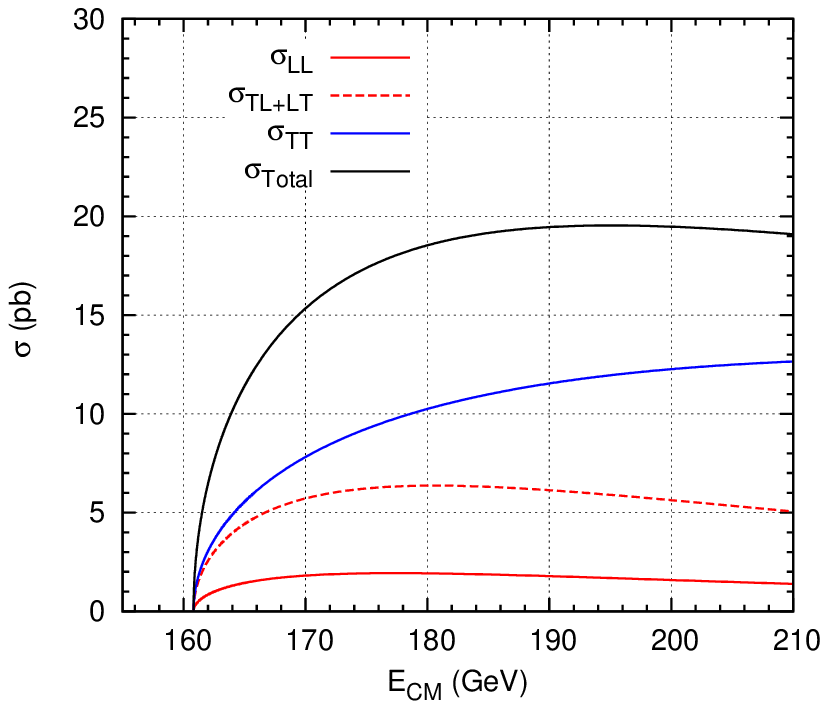}
& \includegraphics[bb=0 60 567 624,width=0.45\linewidth]{wwxsec.eps}
  \end{tabular}
   \caption{Cross section $e^- + e^+ \rightarrow W^- +W^+
     $ for various polarizations of the final $W$ bosons. 
     See text for details.}
\label{fig:wwxs-Pol}
  \end{figure}

\begin{figure}[htb]
  \centering
  \begin{tabular}{cc}
\includegraphics[width=0.45\linewidth]{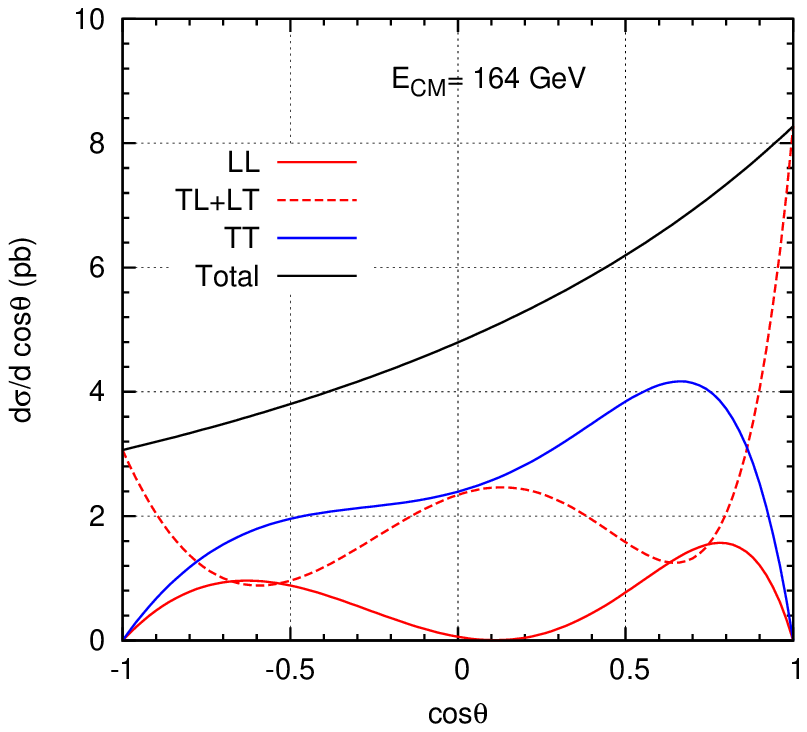}
& \includegraphics[width=0.45\linewidth]{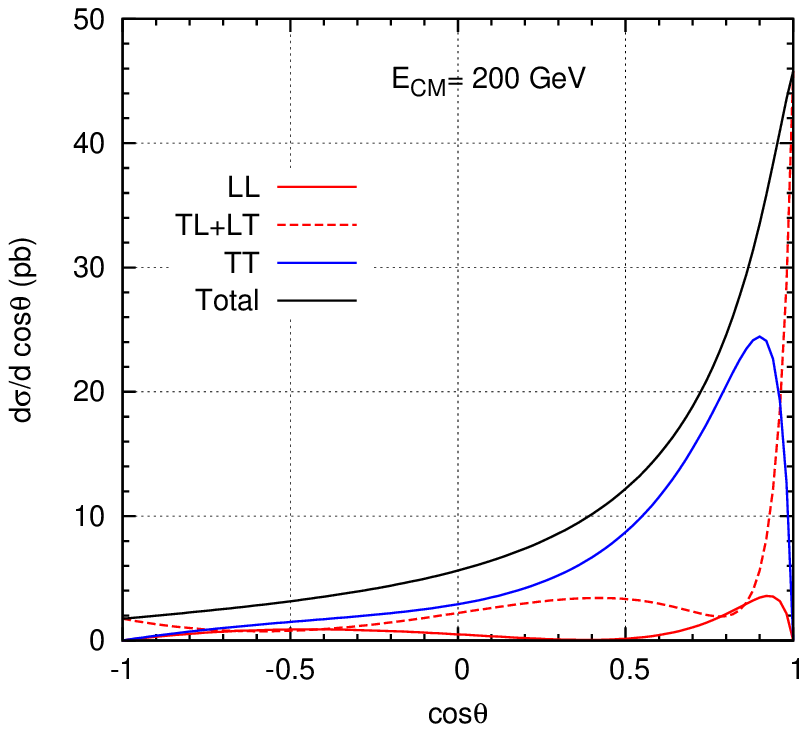}
  \end{tabular}
  \caption{Angular dependence of the differential cross section for
    two values of the center of mass energy.}
\label{fig:difxs}
\end{figure}

As a last exercise let us calculate the cross sections for various
final polarizations of the $W$ bosons. First, let us look at the total
cross sections for the various possibilities. This is shown in
Fig.~\ref{fig:wwxs-Pol}, confirming what we said before concerning the
relative importance of the various polarizations. On the right panel
of that figure we show LEP results for comparison.
One can also learn something from the angular distributions of the
differential cross sections. These are shown in Fig.~\ref{fig:difxs}
for two values of the center of mass energy, the first one close to
threshold, the second one close to the maximum of the cross section. 
The more relevant fact is that the cross section is peaked in the
forward direction and this feature increases with increasing beam
energy. The other relevant piece of information has to do with the fact
that for $\theta=0$, meaning the $W^-$ in the same direction as the
incident electron, both the \texttt{TT} and \texttt{LL} differential
cross section vanish, while the \texttt{TL+LT} has a maximum. This can
be understood as follows. Due to the (V-A) nature of the charged
current interaction, the electron will have helicity $-1/2$ while the
positron helicity $+1/2$, resulting in a total helicity $-1$ in the
direction of the electron. So if the $W$ bosons are produced in the
same direction their total helicity must be $-1$ and that can only be
achieved if one is transverse and the other longitudinal polarization,
more precisely,
\begin{equation}
  \label{eq:43}
  W^-(\lambda =-1), W^+(\lambda =0),\quad \text{or}\quad
  W^+(\lambda =+1), W^-(\lambda =0)
\end{equation}

\subsection{Precision and speed: \texttt{Mathematica}, \texttt{FORM}
  and \texttt{Fortran}}

\begin{figure}[htb]
  \centering
  \begin{tabular}{cc}
\includegraphics[width=0.45\linewidth]{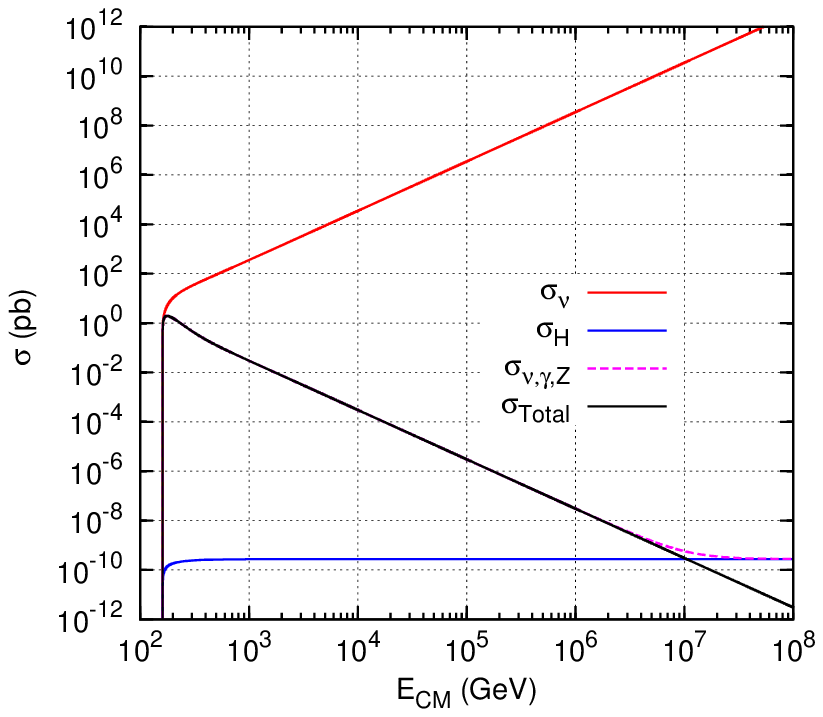}
&
\includegraphics[width=0.45\linewidth]{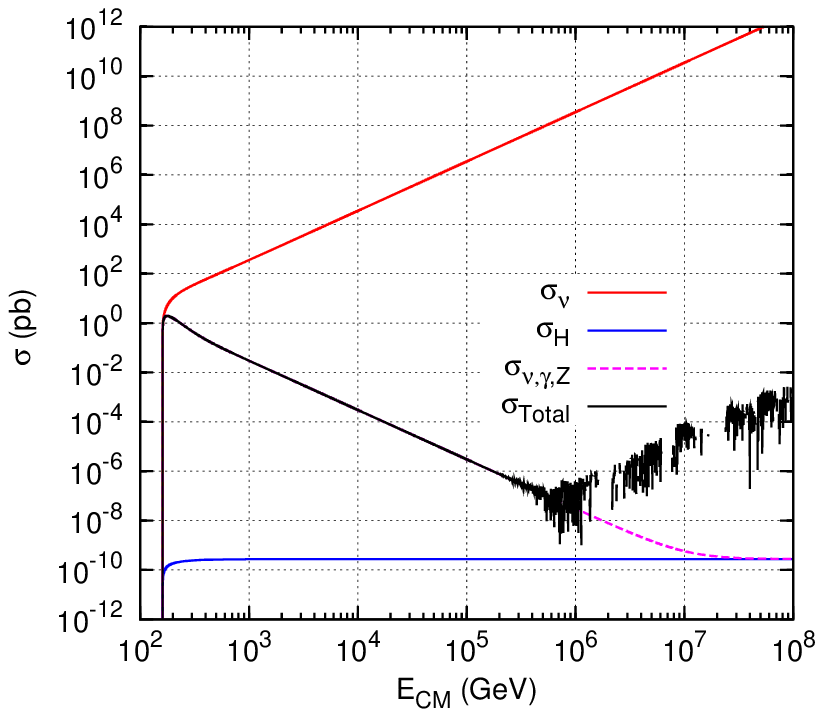} 
  \end{tabular}
  \caption{left panel:
Same as in Fig.~\ref{fig:wwxs-HighEnergy} but with a larger
  scale.}
\label{fig:wwxs-HighEnergy-2}  
\end{figure}

In the previous discussion the more experienced reader should have
been a bit puzzled. To make the point, let us enlarge the plot of
Fig.~\ref{fig:wwxs-HighEnergy}. This is shown in the left panel of
Fig.~\ref{fig:wwxs-HighEnergy-2}. 
We see that for $\sqrt{s}=10^7$ GeV the cancellation has to be
achieved better than one part in $10^{20}$. For numerical calculations
in \texttt{double precision} in \texttt{C} or \texttt{Fortran} this is
a problem. How come that we could achieve this precision without
resorting to \texttt{quadrupole precision}?. The key to the answer
lies in the fact that our formula for the total cross section already
has the bad behaviour cancelled \textit{before} we insert it into the
\texttt{Fortran} program. This is achieved with
\texttt{Mathematica}. One could think that this makes
\texttt{Mathematica} a better choice to make the calculation of
$|\mathcal{M}|^2$. However \texttt{Mathematica} is quite slow when
compared with, for instance, \texttt{FORM}. For this problem in an
\texttt{Intel Core-2} at 2.56 MHz, it takes close to 350 s. The same
problem with \texttt{FORM} takes less than 7 s, a factor of 50!. For
larger problems, one has to use \texttt{FORM}. The problem with
\texttt{FORM} is that has poor capabilities for simplifying
expressions. So with \texttt{FORM} we can get (in 7 s) a
\texttt{Fortran} output that is the \textit{sum} of the all the
contributions to $|\mathcal{M}|^2$ without great simplification. If we
use this output we get the situation in the right panel of
Fig.~\ref{fig:wwxs-HighEnergy-2}. The precision problems appear
exactly where they should, at 1 part in $10^{13}$. But we can get the
best of both programs, we can use \texttt{FORM} to evaluate the traces
and input it into \texttt{Mathematica} to simplify the expressions and
make a \texttt{Fortran} output. In this way we can get the same result
as with \texttt{Mathematica} with a total time of around 20 s instead
of 350 s! The whole process can be automatized~\cite{ctqft:2007}.

\section{Conclusions}
We have discussed the importance of the Higgs boson in making the
\texttt{SM} consistent. We have used symbolic calculations with the
package \texttt{FeynCalc} for \texttt{Mathematica}. The programs can
be found at the web page in Ref.\cite{ctqft:2007}.

\section*{Acknowledgments}

This text started as a set of notes that I have prepared for the
IDPASC Schools at Udine (2012) and Braga (2014)\cite{udine:2012}. 
In the end I wrote
the notes and put them in my web page on calculational methods in
quantum field theory~\cite{ctqft:2007}. Since then, several people
asked me where they were published. As the material is generally known
(at least the conclusions) never crossed my mind to publish it as a
regular article. However, hoping that the details can be useful to a
wider audience, I decided now to put them on the arXiv.

\newpage

\appendix

\section{Inputs for \texttt{FeynCalc}}

\subsection{$\nu_e + \overline{\nu}_e \rightarrow W^+_L  +W^-_L $}
\label{ap:nunuWLWL}
We show here the code for the package \texttt{FeynCalc} that was used
in section~\ref{sec:nunuWLWL}.
\begin{center}
\begin{boxedverbatim}
(******************************** Program nunuWLWL.m *************************************
    This program evaluates the amplitudes and cross section for the process

                      nu + nubar -> W^+_L + W^-_L

 Author: Jorge C. Romão
 email: jorge.romao@ist.utl.pt
********************************* Program nunuWLWL.m ************************************)
Remove["Global`*"]

dm[mu_] := DiracMatrix[mu]
dm[5] := DiracMatrix[5]
ds[p_] := DiracSlash[p]
mt[mu_, nu_] := MetricTensor[mu, nu]
fv[p_, mu_] := FourVector[p, mu]
epsilon[a_, b_, c_, d_] := LeviCivita[a, b, c, d]
id[n_] := IdentityMatrix[n]
sp[p_, q_] := ScalarProduct[p, q]
li[mu_] := LorentzIndex[mu]
prop[p_, m_] := ds[p] + m

PVL[Q_,mu_]:= FourVector[Q,mu]

V[a_,b_,mu_,p_,k_,q_]:=mt[a,b] fv[p-k,mu] + mt[b,mu] fv[k-q,a] + \
                       mt[mu,a] fv[q-p,b] 
PL=dm[7]
PR=dm[6]

(* Functions *)
TakeLimit=Function[{exp}, aux1=exp /. highenergy; aux2= aux1 /. x->1/z; \
	   aux3=Normal[Series[aux2,{z,0,0}]]; aux4=Expand[aux3 /. z->1/x]]

CoefA=Function[{exp},Coefficient[TakeLimit[exp],x,2]]
CoefB=Function[{exp},Coefficient[TakeLimit[exp],x,1]]
CoefC=Function[{exp},Coefficient[TakeLimit[exp],x,0]]

WriteMandel=Function[{exp}, TmpAux1 = exp /. onshell; \
TmpAux2=Simplify[TmpAux1 /.ct]; Simplify[TmpAux2 /. beta->Sqrt[1-4 MW^2/s]]] ; 

WriteTeX=Function[{exp}, TmpAux1 = WriteMandel[exp]; TmpAux2=TeXForm[TmpAux1]]

(* incoming: p1=nu, p2=nubar outgoing: q1=W+, q2=W- *)

onshell1={sp[p1,p1]->0,sp[p2,p2]->0,sp[q1,q1]->MW^2,sp[q2,q2]->MW^2}
onshell2={sp[Q1,q1]->0,sp[Q2,q2]->0}
onshell3={sp[Q1,p1]->s/4/MW (beta - cteta),sp[Q1,p2]->s/4/MW (beta + cteta),
sp[Q1,q2]->s/2/MW beta,sp[Q2,p1]->s/4/MW (beta + cteta),
sp[Q2,p2]->s/4/MW (beta -  cteta),sp[Q2,q1]->s/2/MW beta}
onshell4={sp[Q1,Q1]->-1,sp[Q2,Q2]->-1}
onshell5={sp[Q1,Q2]->s/4/MW^2 (beta^2 +1)}
onshell=Flatten[{onshell1,onshell2,onshell3,onshell4,onshell5}]
\end{boxedverbatim}
\end{center}

\begin{center}
\begin{boxedverbatim}
(****************************************************************************************)
mandel={sp[p1,p2]-> s/2,sp[q1,q2]-> (s -2 MW^2)/2, sp[p1,q1]-> (MW^2  -t)/2,
sp[p2,q2]-> ( MW^2-t)/2, sp[p1,q2]-> (s+t- MW^2)/2,sp[p2,q1]-> (s+t- MW^2)/2}

highenergy={s->4 MW^2 x,t-> MW^2 (1 - 2 x (1- Sqrt[(1 -1/x)] cteta)),beta->Sqrt[1-1/x]}
weakangle={sw^2->1 -cw^2}

ct = Part[Solve[t == MW^2 - s/2*(1 - beta*cteta), cteta], 1]

ampsZ= - 1/2 PVL[Q1,mu] PVL[Q2,nu] Spinor[-p2] . dm[a] .  PL . Spinor[p1] 
V[mu,nu,b,-q1,-q2,q1+q2] (-mt[a,b] + fv[p1+p2,a] fv[p1+p2,b]/MW^2)/(s-MW^2/cw^2)

MsZ = Simplify[DiracSimplify[Contract[ampsZ] /. onshell] /. mandel]

ampNeu= - 1/2 Spinor[-p2] . dm[nu] . PL . prop[p1-q1,0] . dm[mu] . PL . Spinor[p1] 
          PVL[Q1,mu] PVL[Q2,nu] 1/t

MtNeu= Simplify[DiracSimplify[Contract[ampNeu] /. onshell] /. mandel]

diracsub1={DiracGamma[Momentum[Q1]]->1/MW DiracGamma[Momentum[q1]],
DiracGamma[Momentum[Q2]]->1/MW DiracGamma[Momentum[q2]]};

diracsub2={Spinor[-Momentum[p2], 0, 1] . (DiracGamma[Momentum[q1]]/MW) .
 DiracGamma[7] . Spinor[Momentum[p1], 0, 1]-> 1/MW
 Spinor[-Momentum[p2], 0, 1] . DiracGamma[Momentum[q1]] . 
 DiracGamma[7] . Spinor[Momentum[p1], 0, 1],Spinor[-Momentum[p2], 0, 1] . 
 (DiracGamma[Momentum[q2]]/MW) . DiracGamma[7] . 
 Spinor[Momentum[p1], 0, 1]-> 1/MW Spinor[-Momentum[p2], 0, 1] . 
 DiracGamma[Momentum[q2]] . DiracGamma[7] . Spinor[Momentum[p1], 0, 1]};

diracsub3={DiracGamma[Momentum[q2]]->DiracGamma[Momentum[p1+p2-q1]]}

MsZaux1= MsZ /. diracsub1
MsZaux2= Simplify[MsZaux1 /. diracsub2]
MsZaux3= MsZaux2 /. highenergy
MsZaux4= MsZaux3 /. x->1/z
MsZaux5= Normal[Series[MsZaux4,{z,0,1}]] 

MtNeuaux1 = DiracSimplify[MtNeu /. diracsub1]
MtNeuaux2 = MtNeuaux1 /. onshell
MtNeuaux3 = MtNeuaux2 /. highenergy
MtNeuaux4 = MtNeuaux3 /. x->1/z
MtNeuaux5 = Normal[Series[MtNeuaux4,{z,0,1}]] 

Mtotal= Simplify[MsZaux5 + MtNeuaux5]

MHighEnergy:= Simplify[DiracSimplify[Mtotal  /. diracsub3] /. z->1/x]

Print["1) High Energy Behaviour of the Amplitudes: \n x=s/(4 MW^2) , 
\[CurlyPhi](-p2).(\[Gamma]\[CenterDot]q1).\[Gamma]^7.\[CurlyPhi](p1) 
goes like s"] 
Print["MHighEnergy = ",MHighEnergy]
\end{boxedverbatim}
\end{center}

\begin{center}
\begin{boxedverbatim}
(****************************************************************************************)
(* Cross Section *)
Print["\n 2) Exact Amplitudes Squared: \n"]

line1= ds[p2] . dm[nu] . PL . ds[p1-q1] . dm[mu] . PL . ds[p1] . dm[mup] . PL . ds[p1-q1] . 
dm[nup] . PL PVL[Q1,mu] PVL[Q1,mup] PVL[Q2,nu] PVL[Q2,nup]

cst1 = g^4/4/t^2
ans1= cst1 Contract[Tr[line1]] /. onshell
Msqt= Simplify[ans1 /. mandel] 

Print["|M_t|^2 = ",Msqt /. ct /. beta -> Sqrt[1 - 4 MW^2/s] // Simplify]

line2= ds[p2] . dm[a] . PL . ds[p1] . dm[b] . PL V[mu,nu,a,-q1,-q2,p1+p2] \
  V[mup,nup,b,-q1,-q2,p1+p2]  PVL[Q1,mu] PVL[Q1,mup] PVL[Q2,nu] PVL[Q2,nup] 

cst2= g^4/4/(s-MZ^2)^2
ans2= cst2 Contract[Tr[line2]] /. onshell
Msqs= Simplify[ans2 /. mandel] 

Print["|M_s|^2 = ",Msqs /. ct /. beta -> Sqrt[1 - 4 MW^2/s] // Simplify]

subeps={Eps[Momentum[a_],Momentum[b_],Momentum[c_],Momentum[d_]]->0}

line12=ds[p2] . dm[nu] . PL . (ds[p1]-ds[q1]) . dm[mu] . PL . ds[p1] . dm[a] . 
PL PVL[Q1,mu] PVL[Q1,mup] PVL[Q2,nu] PVL[Q2,nup] V[mup,nup,a,-q1,-q2,p1+p2]

line21=ds[p2] . dm[a] . PL . ds[p1] . dm[mu] . PL . (ds[p1]-ds[q1]). dm[nu] . 
PL PVL[Q1,mu] PVL[Q1,mup] PVL[Q2,nu] PVL[Q2,nup] V[mup,nup,a,-q1,-q2,p1+p2]

cst12= -g^4/4/t/(s-MZ^2)
ans12=cst12 Contract[Tr[line12]+Tr[line21]] /. onshell
ans12= Simplify[ans12 /. mandel]
Msqstts= Simplify[ans12 /. subeps]
Print["M_s M_t^* + M_s^* M_t = ",Msqstts /. ct /. beta -> Sqrt[1 - 4 MW^2/s] // Simplify]

MsqTotal=Simplify[Msqt+Msqs+Msqstts];

(* High Energy Limit *)
Print["\n 3) High Energy Limit of Amplitudes Squared: \n"]

MsqtHEL=TakeLimit[Msqt] /. {cteta^2 g^4 x^2 -> g^4 x^2 ( 1 - steta^2), 
    cteta^2 g^4 x -> g^4 x (1 - steta^2)} // Simplify // Expand
Print["|M_t|^2 = ",MsqtHEL]

MsqsHEL=TakeLimit[Msqs] /. {cteta^2 g^4 x^2 -> g^4 x^2 ( 1 - steta^2), 
    cteta^2 g^4 x -> g^4 x (1 - steta^2)} // Simplify // Expand
Print["|M_s|^2 = ",MsqsHEL]

MsqsttsHEL=TakeLimit[Msqstts] /. {cteta^2 g^4 x^2 -> g^4 x^2 ( 1 - steta^2), 
    cteta^2 g^4 x -> g^4 x (1 - steta^2)} // Simplify // Expand
Print["2 Re(M_s M_t^*) = ",MsqsttsHEL]

MsqTotalHEL=MsqtHEL + MsqsHEL + MsqsttsHEL
MsqTotalHEL = MsqTotalHEL /. cteta^2->1-steta^2 //Simplify
Print["|M_total|^2 = ",MsqTotalHEL]
\end{boxedverbatim}
\end{center}

\begin{center}
\begin{boxedverbatim}
(****************************************************************************************)
xs=1
If[xs==1,
   stmp=OpenWrite["anstotal.f",FormatType -> FortranForm];
   Write[stmp,MsqTotal /. g^4->32 MW^4 GF^2];
   Close[stmp];
   stmp=OpenWrite["ansneu.f",FormatType -> FortranForm];
   Write[stmp,Msqt /. g^4->32 MW^4 GF^2];
   Close[stmp];
   stmp=OpenWrite["ansZ.f",FormatType -> FortranForm];
   Write[stmp,Msqs /. g^4->32 MW^4 GF^2];
   Close[stmp];
   stmp=OpenWrite["ansneuZ.f",FormatType -> FortranForm];
   Write[stmp,Msqstts /. g^4->32 MW^4 GF^2];
   Close[stmp],Print[" "]];
(****************************************************************************************)
\end{boxedverbatim}
\end{center}

\subsection{$W^-_L + W^+_L  \rightarrow W^-_L  +W^+_L $}
\label{ap:WLWLWLWL}
We show here the code for the package \texttt{FeynCalc} that was used
in section~\ref{sec:WLWLWLWL}.
\begin{center}
\begin{boxedverbatim}
(******************************** Program WLWLWLWL.m *************************************
*    This program evaluates the amplitudes and cross section for the process             *
*                                                                                        *
*                      W^-_L + W^+_L -> W^+_L + W^-_L                                    *
*                                                                                        *
* Author: Jorge C. Romão                                                                 *
* email: jorge.romao@ist.utl.pt                                                          *
*****************************************************************************************)
Remove["Global`*"]

dm[mu_] := DiracMatrix[mu]
dm[5] := DiracMatrix[5]
ds[p_] := DiracSlash[p]
mt[mu_, nu_] := MetricTensor[mu, nu]
fv[p_, mu_] := FourVector[p, mu]
epsilon[a_, b_, c_, d_] := LeviCivita[a, b, c, d]
id[n_] := IdentityMatrix[n]
sp[p_, q_] := ScalarProduct[p, q]
li[mu_] := LorentzIndex[mu]
prop[p_, m_] := ds[p] + m

PVL[Q_,mu_]:= FourVector[Q,mu]
V[a_,b_,mu_,p_,k_,q_]:=mt[a,b] fv[p-k,mu] + mt[b,mu] fv[k-q,a] + \
                       mt[mu,a] fv[q-p,b] 

PL:=(1 - dm[5])/2
PR:=(1 + dm[5])/2

(****************************************************************************************)
\end{boxedverbatim}
\end{center}

\begin{center}
\begin{boxedverbatim}
(* incoming: p1=W-, p2=W+ outgoing: q1=W-, q2=W+ *)

(* definitions *)
os1={sp[p1,p1]->MW^2,sp[p2,p2]->MW^2,sp[q1,q1]->MW^2,sp[q2,q2]->MW^2}
os2={sp[P1,p1]->0,sp[P2,p2]->0,sp[Q1,q1]->0,sp[Q2,q2]->0}
os3={sp[P1,p2]->s/2/MW beta,sp[P1,q1]->s/4/MW beta (1-cteta)}
os4={sp[P1,q2]->s/4/MW beta (1+cteta),sp[P2,p1]->s/2/MW beta}
os5={sp[P2,q1]->s/4/MW beta (1+cteta),sp[P2,q2]->s/4/MW beta (1-cteta)}
os6={sp[Q1,p1]->s/4/MW beta (1-cteta),sp[Q1,p2]->s/4/MW beta (1+cteta)}
os7={sp[Q1,q2]->s/2/MW beta,sp[Q2,p1]->s/4/MW beta (1+cteta)}
os8={sp[Q2,p2]->s/4/MW beta (1-cteta),sp[Q2,q1]->s/2/MW beta}
os9={sp[P1,P1]->-1,sp[P2,P2]->-1,sp[Q1,Q1]->-1,sp[Q2,Q2]->-1}
os10={sp[P1,P2]->s/4/MW^2 (beta^2 +1),sp[P1,Q1]->s/4/MW^2 (beta^2 -cteta)}
os11={sp[P1,Q2]->s/4/MW^2 (beta^2 +cteta), sp[P2,Q1]->s/4/MW^2 (beta^2+cteta)}
os12={sp[P2,Q2]->s/4/MW^2 (beta^2 -cteta), sp[Q1,Q2]->s/4/MW^2 (beta^2 +1)}
onshell=Flatten[{os1,os2,os3,os4,os5,os6,os7,os8,os9,os10,os11,os12}]

m1={sp[p1,p2]-> (s -2 MW^2)/2,sp[q1,q2]-> (s -2 MW^2)/2, sp[p1,q1]-> (2 MW^2-t)/2}
m2={sp[p2,q2]-> (2 MW^2-t)/2, sp[p1,q2]-> (s+t-2 MW^2)/2,sp[p2,q1]-> (s+t-2 MW^2)/2}
mandel=Flatten[{m1,m2}]

highenergy={s->4 MW^2 x,t->2 MW^2 (1 - x (1- (1 -1/x) cteta)),beta->Sqrt[1-1/x]}
weakangle={sw->Sqrt[1 -cw^2]}
ct = Part[Solve[t == 2 MW^2 - s/2 (1 - beta^2 cteta), cteta], 1]

(* Functions *)
TakeLimit=Function[{exp}, aux1=exp /. highenergy; aux2= aux1 /. x->1/z; \
	   aux3=Normal[Series[aux2,{z,0,0}]]; aux4=Expand[aux3 /. z->1/x]]

CoefA=Function[{exp},Coefficient[TakeLimit[exp],x,2]]
CoefB=Function[{exp},Coefficient[TakeLimit[exp],x,1]]
CoefC=Function[{exp},Coefficient[TakeLimit[exp],x,0]]

WriteMandel=Function[{exp}, TmpAux1 = exp /. onshell; TmpAux2 = Simplify[TmpAux1] ; \
TmpAux3 = Simplify[TmpAux2 /. ct];   TmpAux4=Simplify[TmpAux3 /. beta->Sqrt[1-4 MW^2/s]]]

WriteTeX=Function[{exp}, TmpAux1 = WriteMandel[exp]; TmpAux2=TeXForm[TmpAux1]]

(* s-channel gamma *)
ampsG= sw^2 PVL[P1,a] PVL[P2,b] PVL[Q1,c] PVL[Q2,d] mt[mu,nu] \
       V[a,b,mu,p1,p2,-p1-p2] V[d,c,nu,-q2,-q1,p1+p2] /s

MsG = Simplify[Simplify[Contract[ampsG] /. onshell] /. mandel]
MsG = Simplify[MsG /. weakangle]
MsGA:=CoefA[MsG]
MsGB:=CoefB[MsG]
MsGC:=CoefC[MsG]

(* s-channel Z *)
ampsZ= cw^2 PVL[P1,a] PVL[P2,b] PVL[Q1,c] PVL[Q2,d] \
       V[a,b,mu,p1,p2,-p1-p2] V[d,c,nu,-q2,-q1,p1+p2] \
       (mt[mu,nu] -fv[p1+p2,mu] fv[p1+p2,nu]/MW^2/cw^2)/(s -MW^2/cw^2)

MsZ = Simplify[Simplify[Contract[ampsZ] /. onshell] /. mandel]
MsZ = Simplify[MsZ /. weakangle]
MsZA:=CoefA[MsZ]
MsZB:=CoefB[MsZ]
MsZC:=CoefC[MsZ]
*****************************************************************************************)
\end{boxedverbatim}
\end{center}

\begin{center}
\begin{boxedverbatim}
(* 4 Boson Interaction *)
amp4V = PVL[P1,a] PVL[P2,b] PVL[Q1,c] PVL[Q2,d] \
        (2 mt[a,d] mt[b,c] - mt[a,b] mt[d,c] - mt[a,c] mt[d,b])

M4V = Simplify[Simplify[Contract[amp4V] /. onshell] /. mandel]
M4V = Simplify[M4V /. weakangle]
M4VA:=CoefA[M4V]
M4VB:=CoefB[M4V]
M4VC:=CoefC[M4V]

(* t- channel gamma *)
amptG =  sw^2 PVL[P1,a] PVL[P2,b] PVL[Q1,c] PVL[Q2,d] mt[mu,nu] \
       V[a,c,mu,p1,-q1,q1-p1] V[d,b,nu,-q2,p2,q2-p2] / t

MtG = Simplify[Simplify[Contract[amptG] /. onshell] /. mandel]
MtG = Simplify[MtG /. weakangle]
MtGA:=CoefA[MtG]
MtGB:=CoefB[MtG]
MtGC:=CoefC[MtG]

(* t-channel Z *)
amptZ =  cw^2 PVL[P1,a] PVL[P2,b] PVL[Q1,c] PVL[Q2,d] \
       V[a,c,mu,p1,-q1,q1-p1] V[d,b,nu,-q2,p2,q2-p2] \
       (mt[mu,nu] - fv[p1-q1,mu] fv[p1-q1,nu]/MW^2/cw^2)/(t -MW^2/cw^2)

MtZ = Simplify[Simplify[Contract[amptZ] /. onshell] /. mandel]
MtZ = Simplify[MtZ /. weakangle]
MtZA:=CoefA[MtZ]
MtZB:=CoefB[MtZ]
MtZC:=CoefC[MtZ]
         
(* s-channel Higgs *)
ampsH = - MW^2 PVL[P1,a] PVL[P2,b] PVL[Q1,c] PVL[Q2,d] \
          mt[a,b] mt[c,d] /(s-MH^2)

MsH = Simplify[Simplify[Contract[ampsH] /. onshell] /. mandel]
MsH = Simplify[MsH /. weakangle]
MsHA:=CoefA[MsH]
MsHB:=CoefB[MsH]
MsHC:=CoefC[MsH]

(* t-channel Higgs *)
amptH = - MW^2 PVL[P1,a] PVL[P2,b] PVL[Q1,c] PVL[Q2,d] \
          mt[a,c] mt[b,d] /(t-MH^2)

MtH = Simplify[Simplify[Contract[amptH] /. onshell] /. mandel]
MtH = Simplify[MtH /. weakangle]
MtHA:=CoefA[MtH]
MtHB:=CoefB[MtH]
MtHC:=CoefC[MtH]

testA:=Simplify[MsGA+MsZA+MtGA+MtZA+M4VA+MsHA+MtHA /. weakangle]
testB:=Simplify[MsGB+MsZB+MtGB+MtZB+M4VB+MsHB+MtHB /. weakangle]
*****************************************************************************************)
\end{boxedverbatim}
\end{center}

\begin{center}
\begin{boxedverbatim}
(* Output Fortran *)
xs=1
If[xs==1,
   stmp=OpenWrite["MsG.f",FormatType -> FortranForm];
   Write[stmp,g^2 MsG /. g^2->8 MW^2 GF/Sqrt[2]];
   Close[stmp];
   stmp=OpenWrite["MsZ.f",FormatType -> FortranForm];
   Write[stmp,g^2 MsZ /. g^2->8 MW^2 GF/Sqrt[2]];
   Close[stmp];
   stmp=OpenWrite["MtG.f",FormatType -> FortranForm];
   Write[stmp,g^2 MtG /. g^2->8 MW^2 GF/Sqrt[2]];
   Close[stmp];
   stmp=OpenWrite["MtZ.f",FormatType -> FortranForm];
   Write[stmp,g^2 MtZ /. g^2->8 MW^2 GF/Sqrt[2]];
   Close[stmp];
   stmp=OpenWrite["M4V.f",FormatType -> FortranForm];
   Write[stmp,g^2 M4V /. g^2->8 MW^2 GF/Sqrt[2]];
   Close[stmp];
   stmp=OpenWrite["MsH.f",FormatType -> FortranForm];
   Write[stmp,g^2 MsH /. g^2->8 MW^2 GF/Sqrt[2]];
   Close[stmp];
   stmp=OpenWrite["MtH.f",FormatType -> FortranForm];
   Write[stmp,g^2 MtH /. g^2->8 MW^2 GF/Sqrt[2]];
   Close[stmp];
   ,Print[" "]];
(****************************************************************************************)
\end{boxedverbatim}
\end{center}

\subsection{$e^- + e^+ \rightarrow W^-_L  +W^+_L $}
\label{ap:eEWLWL}

We show here the code for the package \texttt{FeynCalc} that was used
in section~\ref{sec:eEWLWL}.

\begin{center}
\begin{boxedverbatim}
(********************************* Program eEWLWL.m **************************************
*    This program evaluates the amplitudes and cross section for the process             *
*                                                                                        *
*                      e^- + e^+ -> W^+_L + W^-_L                                        *
*                                                                                        *
* Author: Jorge C. Romão                                                                 *
* email: jorge.romao@ist.utl.pt                                                          *
*****************************************************************************************)
Remove["Global`*"] 

(* definitions *)
xs=0
dm[mu_] := DiracMatrix[mu]
dm[5] := DiracMatrix[5]
ds[p_] := DiracSlash[p]
mt[mu_, nu_] := MetricTensor[mu, nu]
fv[p_, mu_] := FourVector[p, mu]
epsilon[a_, b_, c_, d_] := LeviCivita[a, b, c, d]
id[n_] := IdentityMatrix[n]
sp[p_, q_] := Pair[Momentum[p], Momentum[q]]
li[mu_] := LorentzIndex[mu]
prop[p_, m_] := ds[p] + m

pv[p_,mu_]:= PolarizationVector[p,mu]
V[a_,b_,mu_,p_,k_,q_]:=mt[a,b] fv[p-k,mu] + mt[b,mu] fv[k-q,a] + \
                       mt[mu,a] fv[q-p,b] 
\end{boxedverbatim}
\end{center}

\begin{center}
\begin{boxedverbatim}
(* Couplings *)
PL=dm[7]
PR=dm[6]

gR=sw^2
gL=-1/2+sw^2

(* incoming: p1=e-, p2=e+ outgoing: q1=W-, q2=W+ *)
(* kinematics *)
vp1={Sqrt[s]/2,0,0,betae Sqrt[s]/2}
vp2={Sqrt[s]/2,0,0,-betae Sqrt[s]/2}
vq1={Sqrt[s]/2,Sqrt[s]/2 beta steta,0,Sqrt[s]/2 beta cteta}
vq2={Sqrt[s]/2,-Sqrt[s]/2 beta steta,0,-Sqrt[s]/2 beta cteta}
vQ1L={Sqrt[s]/2/MW beta,Sqrt[s]/2/MW  steta,0,Sqrt[s]/2/MW cteta}
vQ2L={Sqrt[s]/2/MW beta,-Sqrt[s]/2/MW  steta,0,-Sqrt[s]/2/MW cteta}
vQ1P={0,cteta/Sqrt[2], I/Sqrt[2] , -steta/Sqrt[2]}
vQ1M={0,cteta/Sqrt[2], -I/Sqrt[2] , -steta/Sqrt[2]}
vQ2P={0,-cteta/Sqrt[2], I/Sqrt[2] , steta/Sqrt[2]}
vQ2M={0,-cteta/Sqrt[2], -I/Sqrt[2] , steta/Sqrt[2]}
vQ1cL=vQ1L
vQ2cL=vQ2L
vQ1cP={0,cteta/Sqrt[2], -I/Sqrt[2] , -steta/Sqrt[2]}
vQ1cM={0,cteta/Sqrt[2], I/Sqrt[2] , -steta/Sqrt[2]}
vQ2cP={0,-cteta/Sqrt[2],-I/Sqrt[2] , steta/Sqrt[2]}
vQ2cM={0,-cteta/Sqrt[2], I/Sqrt[2] , steta/Sqrt[2]}

GetEps=Function[{a,b,c,d},mat={a,b,c,d}; Det[mat]]
GetDot=Function[{a,b},a[[1]]*b[[1]]-a[[2]]*b[[2]]-a[[3]]*b[[3]]-a[[4]]*b[[4]]]

vlist={p1,p2,q1,q2,Q1L,Q2L,Q1P,Q1M,Q2P,Q2M,Q1cL,Q2cL,Q1cP,Q1cM,Q2cP,Q2cM}
fvlist={vp1,vp2,vq1,vq2,vQ1L,vQ2L,vQ1P,vQ1M,vQ2P,vQ2M,vQ1cL,vQ2cL,vQ1cP,vQ1cM,vQ2cP,vQ2cM}
m2list={me^2,me^2,MW^2,MW^2,-1,-1,0,0,0,0,-1,-1,0,0,0,0}
onshell=Flatten[Table[If[i==j,Pair[Momentum[vlist[[i]]],Momentum[vlist[[j]]]]->m2list[[i]],\
Pair[Momentum[vlist[[i]]],Momentum[vlist[[j]]]]->GetDot[fvlist[[i]],fvlist[[j]]]],\
{i,1,16},{j,i,16}]];
onshell = Simplify[onshell /. {steta^2->1-cteta^2,beta^2->1-4 MW^2/s}];

Q1=Polarization[q1];
Q2=Polarization[q2];
Q1c=Momentum[Polarization[q1, -I]];
Q2c=Momentum[Polarization[q2, -I]];

VList={p1,p2,q1,q2,Q1,Q1c,Q2,Q2c}
VListLL={p1,p2,q1,q2,Q1L,Q1cL,Q2L,Q2cL}
VListLP={p1,p2,q1,q2,Q1L,Q1cL,Q2P,Q2cP}
VListLM={p1,p2,q1,q2,Q1L,Q1cL,Q2M,Q2cM}
VListPL={p1,p2,q1,q2,Q1P,Q1cP,Q2L,Q2cL}
VListML={p1,p2,q1,q2,Q1M,Q1cM,Q2L,Q2cL}
VListPP={p1,p2,q1,q2,Q1P,Q1cP,Q2P,Q2cP}
VListPM={p1,p2,q1,q2,Q1P,Q1cP,Q2M,Q2cM}
VListMP={p1,p2,q1,q2,Q1M,Q1cM,Q2P,Q2cP}
VListMM={p1,p2,q1,q2,Q1M,Q1cM,Q2M,Q2cM}

M2List={me^2,me^2,MW^2,MW^2}
ms1=Flatten[Table[sp[VList[[i]],VList[[i]]]->M2List[[i]],{i,1,4}]]
ms2=Flatten[Table[sp[vlist[[i]],vlist[[j]]]-> GetDot[fvlist[[i]],\
fvlist[[j]]],{i,1,4},{j,i+1,4}]]
mshell=ms1
(****************************************************************************************)
\end{boxedverbatim}
\end{center}

\begin{center}
\begin{boxedverbatim}
PolLL = Flatten[{mshell,Table[sp[VList[[i]],VList[[j]]]-> (sp[VListLL[[i]],VListLL[[j]]]\
 /. onshell),{i,1,8},{j,i+1,8}]}];
PolLL = DeleteCases[DeleteCases[PolLL,0->0],-1->-1];
PolLP = Flatten[{mshell,Table[sp[VList[[i]],VList[[j]]]-> (sp[VListLP[[i]],VListLP[[j]]]\
 /. onshell),{i,1,8},{j,i+1,8}]}];
PolLP = DeleteCases[DeleteCases[PolLP,0->0],-1->-1];
PolLM = Flatten[{mshell,Table[sp[VList[[i]],VList[[j]]]-> (sp[VListLM[[i]],VListLM[[j]]]\
 /. onshell),{i,1,8},{j,i+1,8}]}];
PolLM = DeleteCases[DeleteCases[PolLM,0->0],-1->-1];
PolPL = Flatten[{mshell,Table[sp[VList[[i]],VList[[j]]]-> (sp[VListPL[[i]],VListPL[[j]]]\
 /. onshell),{i,1,8},{j,i+1,8}]}];
PolPL = DeleteCases[DeleteCases[PolPL,0->0],-1->-1];
PolML = Flatten[{mshell,Table[sp[VList[[i]],VList[[j]]]-> (sp[VListML[[i]],VListML[[j]]]\
 /. onshell),{i,1,8},{j,i+1,8}]}];
PolML = DeleteCases[DeleteCases[PolML,0->0],-1->-1];
PolPP = Flatten[{mshell,Table[sp[VList[[i]],VList[[j]]]-> (sp[VListPP[[i]],VListPP[[j]]]\
 /. onshell),{i,1,8},{j,i+1,8}]}];
PolPP = DeleteCases[DeleteCases[PolPP,0->0],-1->-1];
PolPM = Flatten[{mshell,Table[sp[VList[[i]],VList[[j]]]-> (sp[VListPM[[i]],VListPM[[j]]]\
 /. onshell),{i,1,8},{j,i+1,8}]}];
PolPM = DeleteCases[DeleteCases[PolPM,0->0],-1->-1];
PolMP = Flatten[{mshell,Table[sp[VList[[i]],VList[[j]]]-> (sp[VListMP[[i]],VListMP[[j]]]\
 /. onshell),{i,1,8},{j,i+1,8}]}];
PolMP = DeleteCases[DeleteCases[PolMP,0->0],-1->-1];
PolMM = Flatten[{mshell,Table[sp[VList[[i]],VList[[j]]]-> (sp[VListMM[[i]],VListMM[[j]]]\
 /. onshell),{i,1,8},{j,i+1,8}]}];
PolMM = DeleteCases[DeleteCases[PolMM,0->0],-1->-1];

fvListLL={vp1,vp2,vq1,vq2,vQ1L,vQ1cL,vQ2L,vQ2cL}
fvListLP={vp1,vp2,vq1,vq2,vQ1L,vQ1cL,vQ2P,vQ2cP}
fvListLM={vp1,vp2,vq1,vq2,vQ1L,vQ1cL,vQ2M,vQ2cM}
fvListPL={vp1,vp2,vq1,vq2,vQ1P,vQ1cP,vQ2L,vQ2cL}
fvListML={vp1,vp2,vq1,vq2,vQ1M,vQ1cM,vQ2L,vQ2cL}
fvListPP={vp1,vp2,vq1,vq2,vQ1P,vQ1cP,vQ2P,vQ2cP}
fvListPM={vp1,vp2,vq1,vq2,vQ1P,vQ1cP,vQ2M,vQ2cM}
fvListMP={vp1,vp2,vq1,vq2,vQ1M,vQ1cM,vQ2P,vQ2cP}
fvListMM={vp1,vp2,vq1,vq2,vQ1M,vQ1cM,vQ2M,vQ2cM}

subepsLL = DeleteCases[Flatten[Table[If[i != j && i != k && i != m && j != k\
	&& j != m && k != m,Eps[Momentum[VList[[i]]], Momentum[VList[[j]]], \
      Momentum[VList[[k]]], Momentum[VList[[m]]]] -> (GetEps[fvListLL[[i]], \
    fvListLL[[j]], fvListLL[[k]], fvListLL[[m]]] /. onshell),1], {i, 1, 8}, \
    {j, 1, 8}, {k, 1, 8}, {m, 1, 8}]],_Integer];

subepsLP = DeleteCases[Flatten[Table[If[i != j && i != k && i != m && j != k \
         && j != m && k != m,Eps[Momentum[VList[[i]]], Momentum[VList[[j]]], \
    Momentum[VList[[k]]], Momentum[VList[[m]]]] -> (GetEps[fvListLP[[i]], \
    fvListLP[[j]], fvListLP[[k]], fvListLP[[m]]] /. onshell),1], {i, 1, 8}, \
    {j, 1, 8},{k, 1, 8}, {m, 1, 8}]],_Integer];

subepsLM = DeleteCases[Flatten[Table[If[i != j && i != k && i != m && j != k \
         && j != m && k != m,Eps[Momentum[VList[[i]]], Momentum[VList[[j]]], \
      Momentum[VList[[k]]], Momentum[VList[[m]]]] -> (GetEps[fvListLM[[i]], \
    fvListLM[[j]], fvListLM[[k]], fvListLM[[m]]] /. onshell),1], {i, 1, 8}, \
    {j, 1, 8},{k, 1, 8}, {m, 1, 8}]],_Integer];

subepsPL = DeleteCases[Flatten[Table[If[i != j && i != k && i != m && j != k \
         && j != m && k != m,Eps[Momentum[VList[[i]]], Momentum[VList[[j]]], \
       Momentum[VList[[k]]], Momentum[VList[[m]]]] -> (GetEps[fvListPL[[i]], \
    fvListPL[[j]], fvListPL[[k]], fvListPL[[m]]] /. onshell),1], {i, 1, 8}, \
    {j, 1, 8},{k, 1, 8}, {m, 1, 8}]],_Integer];
(****************************************************************************************)
\end{boxedverbatim}
\end{center}

\begin{center}
\begin{boxedverbatim}
subepsML = DeleteCases[Flatten[Table[If[i != j && i != k && i != m && j != k \
       && j != m && k != m,Eps[Momentum[VList[[i]]], Momentum[VList[[j]]], \
      Momentum[VList[[k]]], Momentum[VList[[m]]]] -> (GetEps[fvListML[[i]], \
    fvListML[[j]], fvListML[[k]], fvListML[[m]]] /. onshell),1], {i, 1, 8}, \
    {j, 1, 8},{k, 1, 8}, {m, 1, 8}]],_Integer];

subepsPP = DeleteCases[Flatten[Table[If[i != j && i != k && i != m && j!= k \ 
    && j != m && k != m,Eps[Momentum[VList[[i]]], Momentum[VList[[j]]], \
    Momentum[VList[[k]]], Momentum[VList[[m]]]] -> (GetEps[fvListPP[[i]], \
    fvListPP[[j]], fvListPP[[k]], fvListPP[[m]]] /. onshell),1], {i, 1, 8}, \
    {j, 1, 8},{k, 1, 8}, {m, 1, 8}]],_Integer];
subepsPM =DeleteCases[ Flatten[Table[If[i != j && i != k && i != m && j != k \
         && j != m && k != m,Eps[Momentum[VList[[i]]], Momentum[VList[[j]]], \
    Momentum[VList[[k]]], Momentum[VList[[m]]]] -> (GetEps[fvListPM[[i]], \
    fvListPM[[j]], fvListPM[[k]], fvListPM[[m]]] /. onshell),1], {i, 1, 8}, \
    {j, 1, 8}, {k, 1, 8}, {m, 1, 8}]],_Integer];

subepsMP = DeleteCases[Flatten[Table[If[i != j && i != k && i != m && j != k \
         && j != m && k != m,Eps[Momentum[VList[[i]]], Momentum[VList[[j]]], \
    Momentum[VList[[k]]], Momentum[VList[[m]]]] -> (GetEps[fvListMP[[i]], \
    fvListMP[[j]], fvListMP[[k]], fvListMP[[m]]] /. onshell),1], {i, 1, 8}, \
    {j, 1, 8},{k, 1, 8}, {m, 1, 8}]],_Integer];

subepsMM = DeleteCases[Flatten[Table[If[i != j && i != k && i != m && j != k \
         && j != m && k != m,Eps[Momentum[VList[[i]]], Momentum[VList[[j]]], \
    Momentum[VList[[k]]], Momentum[VList[[m]]]] -> (GetEps[fvListMM[[i]], \
    fvListMM[[j]], fvListMM[[k]], fvListMM[[m]]] /. onshell),1], {i, 1, 8}, \
    {j, 1, 8}, {k, 1, 8}, {m, 1, 8}]],_Integer];

Pol={PolLL,PolLM,PolLP,PolML,PolPL,PolPP,PolPM,PolMP,PolMM}
subeps={subepsLL,subepsLM,subepsLP,subepsML,subepsPL,subepsPP,subepsPM,subepsMP,subepsMM}
simpbetas={beta -> Sqrt[1 - 4 MW^2/s], betae -> Sqrt[1 - 4 me^2/s]}
weakangle={sw->Sqrt[1 -cw^2]}

he1={s->4 MW^2 x}
he2={t-> MW^2 + me^2 - 2 MW^2 x (1- Sqrt[(1 -1/x)] Sqrt[1-me^2/MW^2/x] cteta)}
he3={beta->Sqrt[1-1/x]}
he4={betae->Sqrt[1-me^2/x/MW^2]}
highenergy=Flatten[{he1,he2,he3,he4}]

weakangle={sw->Sqrt[1 -cw^2]}
simpbetas={beta -> Sqrt[1 - 4 MW^2/s], betae -> Sqrt[1 - 4 me^2/s]}

ds1LL={DiracGamma[Momentum[Polarization[q1,I]]]->1/MW DiracGamma[Momentum[q1]]};
ds2LL={DiracGamma[Momentum[Polarization[q2,I]]]->1/MW DiracGamma[Momentum[q2]]};
diracsubLL=Flatten[{ds1LL,ds2LL}];

diracsubLT={DiracGamma[Momentum[Polarization[q1, I]]]->1/MW DiracGamma[Momentum[q1]]};
diracsubTL={DiracGamma[Momentum[Polarization[q2, I]]]->1/MW DiracGamma[Momentum[q2]]};
diracsubTT={};

diracsub1=diracsubLL
diracsub2={DiracGamma[Momentum[q2]]->DiracGamma[Momentum[p1+p2-q1]]}

PolType=Pol[[1]]
(****************************************************************************************)
\end{boxedverbatim}
\end{center}

\begin{center}
\begin{boxedverbatim}
(* Functions *)
TakeLimit=Function[{exp,n}, aux1=exp /. highenergy; aux2= aux1 /. x->1/z; \
 aux3=Normal[Series[aux2,{z,0,n}]]; aux4=Simplify[Expand[aux3 /. z->1/x]]]

ct = Part[Solve[t == me^2+ MW^2 - s/2 (1 - beta*betae cteta), cteta], 1]
st2 ={steta^2->1-cteta^2}
beta2={beta^2->1-4 MW^2/s}

WriteMandel=Function[{exp}, TmpAux1 = exp /. onshell; TmpAux2 = Simplify[TmpAux1 /. Pol] ; \
TmpAux3 = Simplify[TmpAux2 /. ct];   TmpAux4=Simplify[TmpAux3 /. beta->Sqrt[1-4 MW^2/s]]]
WriteTeX=Function[{exp}, TmpAux1 = WriteMandel[exp]; TmpAux2=TeXForm[TmpAux1]]

(* Amplitudes *)
LineG= Spinor[-p2,me] . dm[a] . (PL + PR) . Spinor[p1,me] 
ampsG= sw^2 LineG pv[q1,mu] pv[q2,nu] V[nu,mu,b,-q2,-q1,p1+p2] (-mt[a,b])/s
MsG = DiracSimplify[Contract[ampsG] /. onshell] 

LineZ = Spinor[-p2,me] . dm[a] . (gL PL + gR PR).  Spinor[p1,me]
PropZ = (-mt[a,b] +  cw^2 fv[p1+p2,a] fv[p1+p2,b]/MW^2)/(s-MW^2/cw^2)
ampsZ= - pv[q1,mu] pv[q2,nu] LineZ V[nu,mu,b,-q2,-q1,p1+p2] PropZ
MsZ = DiracSimplify[Contract[ampsZ] /. onshell] 

LineNeu=Spinor[-p2,me] . dm[nu] . PL . prop[p1-q1,0] . dm[mu] . PL . Spinor[p1,me]
ampNeu= - 1/2 LineNeu pv[q1,mu] pv[q2,nu] /t
MtNeu= DiracSimplify[Contract[ampNeu] /. onshell] 

LineH = Spinor[-p2,me] . (PL+PR). Spinor[p1,me]
ampH =  me/2 LineH pv[q1,mu] pv[q2,nu] mt[mu,nu]/(s-mH^2)
MsH= DiracSimplify[Contract[ampH] /. onshell]

(* High Energy Limit *)
MtNeuaux= DiracSimplify[MtNeu /. diracsub1]
MtNeuaux= DiracSimplify[MtNeuaux /. diracsub2]
MtNeuaux= DiracSimplify[MtNeuaux /. PolType]
MtNeuaux= DiracSimplify[MtNeuaux /. diracsub1]
MtNeuHEL = TakeLimit[MtNeuaux,0]

MsGaux= DiracSimplify[MsG /. diracsub1]
MsGaux= DiracSimplify[MsGaux /. diracsub2]
MsGaux= DiracSimplify[MsGaux /. PolType]
MsGaux= DiracSimplify[MsGaux /. diracsub1]
MsGHEL = TakeLimit[MsGaux,0]

MsZaux= DiracSimplify[MsZ /. diracsub1]
MsZaux= DiracSimplify[MsZaux /. diracsub2]
MsZaux= DiracSimplify[MsZaux /. PolType]
MsZaux= DiracSimplify[MsZaux /. diracsub1]
MsZHEL = TakeLimit[MsZaux,0]

MsHaux= DiracSimplify[MsH /. diracsub1]
MsHaux= DiracSimplify[MsHaux /. diracsub2]
MsHaux= DiracSimplify[MsHaux /. PolType]
MsHHEL = TakeLimit[MsHaux,0]

MFinalHEL:= Simplify[MtNeuHEL+MsGHEL+MsZHEL+MsHHEL]
(****************************************************************************************)
\end{boxedverbatim}
\end{center}

\begin{center}
\begin{boxedverbatim}
(* cross section *)
GammaNu =  - 1/(2 t) dm[nu] . PL . prop[p1-q1,0] . dm[mu] . PL 
GammaZ1 = - dm[b] . (gL PL + gR PR) V[nu,mu,b,-q2,-q1,p1+p2] 
GammaZ2 = (-me) (gL PL + gR PR) V[nu,mu,b,-q2,-q1,p1+p2] cw^2 fv[p1+p2,b]/MW^2 
GammaZ3 = (me) (gL PR + gR PL) V[nu,mu,b,-q2,-q1,p1+p2]  cw^2 fv[p1+p2,b]/MW^2 
GammaZ  = - 1/(s-MW^2/cw^2) (GammaZ1+GammaZ2+GammaZ3)
GammaG  =  sw^2/s  dm[a] . (PL + PR)  V[nu,mu,b,-q2,-q1,p1+p2] (-mt[a,b])
GammaH  =  me/2 (PL+PR) mt[mu,nu]/(s-mH^2)
GammaBarNu =  - 1/(2 t) dm[mup] . PL . prop[p1-q1,0] . dm[nup] . PL 
GammaBarZ1 = - dm[bp] . (gL PL + gR PR) V[nup,mup,bp,-q2,-q1,p1+p2] 
GammaBarZ2 = (-me) (gL PL + gR PR) V[nup,mup,bp,-q2,-q1,p1+p2] cw^2 fv[p1+p2,bp]/MW^2 
GammaBarZ3 = (me) (gL PR + gR PL) V[nup,mup,bp,-q2,-q1,p1+p2]  cw^2 fv[p1+p2,bp]/MW^2 
GammaBarZ  = - 1/(s-MW^2/cw^2) (GammaBarZ1+GammaBarZ2+GammaBarZ3)
GammaBarG  =  sw^2/s  dm[ap] . (PL + PR)  V[nup,mup,bp,-q2,-q1,p1+p2] (-mt[ap,bp])
GammaBarH  =  me/2 (PL+PR) mt[mup,nup]/(s-mH^2)


(* Neutrino + gamma + Z cross section *)
LineNuNu =  prop[p2,-me] . GammaNu . prop[p1,me] . GammaBarNu
MsquareNuNu=Contract[Tr[LineNuNu] pv[q1,mu] pv[q2,nu] Conjugate[pv[q1,mup]]\
 Conjugate[pv[q2,nup]]] /. mshell;
MsqNuNu = Table[(MsquareNuNu /. subeps[[i]]) /. Pol[[i]] // Simplify,{i,1,9}]
MsqNuNu = (MsqNuNu /. st2) /. beta2 //Simplify
Clear[MsquareNuNu]

LineNuG =  prop[p2,-me] . GammaNu . prop[p1,me] . GammaBarG 
MsquareNuG = Contract[Tr[LineNuG] pv[q1,mu] pv[q2,nu] Conjugate[pv[q1,mup]]\
  Conjugate[pv[q2,nup]]] /. mshell;
MsqNuG = Table[(MsquareNuG /. subeps[[i]]) /. Pol[[i]] // Simplify,{i,1,9}]
MsqNuG = (MsqNuG /. st2) /. beta2 //Simplify
Clear[MsquareNuG]

LineGNu =  prop[p2,-me] . GammaG . prop[p1,me] . GammaBarNu
MsquareGNu := Contract[Tr[LineGNu] pv[q1,mu] pv[q2,nu]Conjugate[pv[q1,mup]]\
 Conjugate[pv[q2,nup]]] /. mshell;
MsqGNuAux:=Table[(MsquareGNu /. subeps[[i]]) /. Pol[[i]] // Simplify,{i,1,9}];
TestGNuNuG=Function[{},Msq=MsquareGNu; MsqGNuAux=Table[(Msq/. subeps[[i]]) \
/. Pol[[i]] // Simplify,{i,1,9}];  Clear[Msq,MsquareGNu]; MsqGNu =(MsqGNuAux \
/. st2) /. beta2 //Simplify; Table[MsqNuG[[i]]-MsqGNu[[i]],{i,1,9}]]

LineNuZ =  prop[p2,-me] . GammaNu . prop[p1,me] . GammaBarZ
MsquareNuZ = Contract[Tr[LineNuZ] pv[q1,mu] pv[q2,nu] Conjugate[pv[q1,mup]]\
  Conjugate[pv[q2,nup]]] /. mshell;
MsqNuZ = Table[(MsquareNuZ /. subeps[[i]]) /. Pol[[i]] // Simplify,{i,1,9}]
MsqNuZ = (MsqNuZ /. st2) /. beta2 //Simplify
Clear[MsquareNuZ]

LineZNu =  prop[p2,-me] . GammaZ . prop[p1,me] . GammaBarNu
MsquareZNu := Contract[Tr[LineZNu] pv[q1,mu] pv[q2,nu] Conjugate[pv[q1,mup]]\
  Conjugate[pv[q2,nup]]] /. mshell;
MsqZNuAux:=Table[(MsquareZNu /. subeps[[i]]) /. Pol[[i]] // Simplify,{i,1,9}];
TestZNuNuZ=Function[{},Msq=MsquareZNu; MsqZNuAux=Table[(Msq /. subeps[[i]]) \
/. Pol[[i]] // Simplify,{i,1,9}];  Clear[Msq,MsquareZNu]; MsqZNu = (MsqZNuAux \
/. st2) /. beta2 //Simplify; Table[MsqNuZ[[i]]-MsqZNu[[i]],{i,1,9}]]

LineGG =  prop[p2,-me] . GammaG . prop[p1,me] . GammaBarG
MsquareGG =  Contract[Tr[LineGG] pv[q1,mu] pv[q2,nu]Conjugate[pv[q1,mup]] \
 Conjugate[pv[q2,nup]]] /. mshell;
MsqGG = Table[(MsquareGG /. subeps[[i]]) /. Pol[[i]] // Simplify,{i,1,9}]
MsqGG = (MsqGG /. st2) /. beta2 //Simplify
Clear[MsquareGG]
(*****************************************************************************************)
\end{boxedverbatim}
\end{center}

\begin{center}
\begin{boxedverbatim}
LineZZ =  prop[p2,-me] . GammaZ . prop[p1,me] . GammaBarZ
MsquareZZ =  Contract[Tr[LineZZ] pv[q1,mu] pv[q2,nu] Conjugate[pv[q1,mup]] \
 Conjugate[pv[q2,nup]]] /. mshell;
MsqZZ = Table[(MsquareZZ /. subeps[[i]]) /. Pol[[i]] // Simplify,{i,1,9}]
MsqZZ = (MsqZZ /. st2) /. beta2 //Simplify
Clear[MsquareZZ]
 
LineGZ =  prop[p2,-me] . GammaG . prop[p1,me] . GammaBarZ
MsquareGZ =  Contract[Tr[LineGZ] pv[q1,mu] pv[q2,nu] Conjugate[pv[q1,mup]] \
 Conjugate[pv[q2,nup]]] /. mshell;
MsqGZ = Table[(MsquareGZ /. subeps[[i]]) /. Pol[[i]] // Simplify,{i,1,9}]
MsqGZ = (MsqGZ /. st2) /. beta2 //Simplify
Clear[MsquareGZ]

LineZG =  prop[p2,-me] . GammaZ . prop[p1,me] . GammaBarG
MsquareZG:=  Contract[Tr[LineZG] pv[q1,mu] pv[q2,nu] Conjugate[pv[q1,mup]] \
 Conjugate[pv[q2,nup]]] /. mshell;
MsqZGAux:=Table[(MsquareZG /. subeps[[i]]) /. Pol[[i]] // Simplify,{i,1,9}];
TestZGGZ=Function[{},Msq=MsquareZG; Clear[MsquareZG]; MsqZGAux=Table[(Msq \
/. subeps[[i]]) /. Pol[[i]] // Simplify,{i,1,9}]; Clear[Msq,MsquareZG]; MsqZG \
= (MsqZGAux /. st2) /. beta2 //Simplify; Table[MsqGZ[[i]]-MsqZG[[i]],{i,1,9}]]

LineHH =  prop[p2,-me] . GammaH . prop[p1,me] . GammaBarH
MsquareHH =  Contract[Tr[LineHH] pv[q1,mu] pv[q2,nu] Conjugate[pv[q1,mup]]\
  Conjugate[pv[q2,nup]]] /. mshell;
MsqHH = Table[(MsquareHH /. subeps[[i]]) /. Pol[[i]] // Simplify,{i,1,9}]
MsqHH = (MsqHH /. st2) /. beta2 //Simplify
Clear[MsquareHH]

LineNuH =  prop[p2,-me] . GammaNu . prop[p1,me] . GammaBarH
MsquareNuH = Contract[Tr[LineNuH] pv[q1,mu] pv[q2,nu] Conjugate[pv[q1,mup]] \
 Conjugate[pv[q2,nup]]] /. mshell;
MsqNuH = Table[(MsquareNuH /. subeps[[i]]) /. Pol[[i]] // Simplify,{i,1,9}]
MsqNuH = (MsqNuH /. st2) /. beta2 //Simplify
Clear[MsquareNuH]

LineHNu =  prop[p2,-me] . GammaH . prop[p1,me] . GammaBarNu
MsquareHNu:= Contract[Tr[LineHNu] pv[q1,mu] pv[q2,nu] Conjugate[pv[q1,mup]] \ 
 Conjugate[pv[q2,nup]]] /. mshell;
MsqHNuAux:=Table[(MsquareHNu /. subeps[[i]]) /. Pol[[i]] // Simplify,{i,1,9}];
TestHNuNuH=Function[{},Msq=MsquareHNu; Clear[MsquareHNu]; MsqHNuAux=Table[(Msq \
/. subeps[[i]]) /. Pol[[i]] // Simplify,{i,1,9}]; Clear[Msq,MsquareHNu]; \
MsqHNu = (MsqHNuAux /. st2) /. beta2 //Simplify; Table[MsqNuH[[i]]-MsqHNu[[i]],{i,1,9}]]

LineGH =  prop[p2,-me] . GammaG . prop[p1,me] . GammaBarH
MsquareGH = Contract[Tr[LineGH] pv[q1,mu] pv[q2,nu] Conjugate[pv[q1,mup]]\
  Conjugate[pv[q2,nup]]] /. mshell;
MsqGH = Table[(MsquareGH /. subeps[[i]]) /. Pol[[i]] // Simplify,{i,1,9}]
MsqGH = (MsqGH /. st2) /. beta2 //Simplify
Clear[MsquareGH]

LineHG =  prop[p2,-me] . GammaH . prop[p1,me] . GammaBarG
MsquareHG:= Contract[Tr[LineHG] pv[q1,mu] pv[q2,nu] Conjugate[pv[q1,mup]] \
 Conjugate[pv[q2,nup]]] /. mshell;
MsqHGAux:=Table[(MsquareHG /. subeps[[i]]) /. Pol[[i]] // Simplify,{i,1,9}];
TestHGGH=Function[{},Msq=MsquareHG; Clear[MsquareHG]; MsqHGAux=Table[(Msq /. \
subeps[[i]]) /. Pol[[i]] // Simplify,{i,1,9}]; Clear[Msq,MsquareHG]; MsqHG =\
 (MsqHGAux /. st2) /. beta2 //Simplify; Table[MsqGH[[i]]-MsqHG[[i]],{i,1,9}]]
(****************************************************************************************)
\end{boxedverbatim}
\end{center}

\begin{center}
\begin{boxedverbatim}
LineZH =  prop[p2,-me] . GammaZ . prop[p1,me] . GammaBarH
MsquareZH = Contract[Tr[LineZH] pv[q1,mu] pv[q2,nu] Conjugate[pv[q1,mup]]\
  Conjugate[pv[q2,nup]]] /. mshell;
MsqZH = Table[(MsquareZH /. subeps[[i]]) /. Pol[[i]] // Simplify,{i,1,9}]
MsqZH = (MsqZH /. st2) /. beta2 //Simplify
Clear[MsquareZH]

LineHZ =  prop[p2,-me] . GammaH . prop[p1,me] . GammaBarZ
MsquareHZ:= Contract[Tr[LineHZ] pv[q1,mu] pv[q2,nu]Conjugate[pv[q1,mup]] \
 Conjugate[pv[q2,nup]]] /. mshell;
MsqHZAux:=Table[(MsquareHZ /. subeps[[i]]) /. Pol[[i]] // Simplify,{i,1,9}];
TestHZZH=Function[{},Msq=MsquareHZ; Clear[MsquareHZ]; MsqHZAux=Table[(Msq \
/. subeps[[i]]) /. Pol[[i]] // Simplify,{i,1,9}]; Clear[Msq,MsquareHZ]; MsqHZ =\
 (MsqHZAux /. st2) /. beta2 //Simplify; Table[MsqZH[[i]]-MsqHZ[[i]],{i,1,9}]]

MsqNuLL = Simplify[MsqNuNu[[1]]*g^4/4 /. ct] /. g^4->32 MW^4 GF^2
MsqNuLL = Simplify[MsqNuLL /. simpbetas];

MsqHLL = Simplify[MsqHH[[1]]*g^4/4 /. ct ] /. g^4->32 MW^4 GF^2 ;
MsqHLL= Simplify[MsqHLL /. simpbetas];

MsqNuGZLL=MsqNuNu[[1]] + MsqGG[[1]] + MsqZZ[[1]] + \
          2*MsqNuG[[1]] + 2*MsqNuZ[[1]] + 2*MsqGZ[[1]];
MsqNuGZLL = Simplify[MsqNuGZLL*g^4/4 /. ct] /. g^4->32 MW^4 GF^2 ;
MsqNuGZLL =Simplify[MsqNuGZLL /. simpbetas];

MsqHiggsLL = MsqHH[[1]] + 2*MsqNuH[[1]] + 2*MsqGH[[1]] + 2*MsqZH[[1]] ;
MsqHiggsLL = Simplify[MsqHiggsLL*g^4/4 /. ct] /. g^4->32 MW^4 GF^2 ;
MsqHiggsLL = Simplify[MsqHiggsLL /. simpbetas];

MsqLL = MsqNuGZLL + MsqHiggsLL ;

MsqLL=Simplify[MsqLL /. steta->Sqrt[1-cteta^2]];
MsqLL=Simplify[MsqLL /. ct];
MsqLL=Simplify[MsqLL /. simpbetas];

If[xs==1,
   SetOptions[$Output, PageWidth -> 60];
   stmp=OpenWrite["ansNu.f",FormatType -> FortranForm];
   Write[stmp,"MsqNu      =      ", MsqNuLL];
   Close[stmp];
   stmp=OpenWrite["ansNuGZ.f",FormatType -> FortranForm];
   Write[stmp,"MsqNuGZ      =      ", MsqNuGZLL];
   Close[stmp];
   stmp=OpenWrite["ansLL.f",FormatType -> FortranForm];
   Write[stmp,"MsqLL      =      ", MsqLL];
   Close[stmp];
   stmp=OpenWrite["ansH.f",FormatType -> FortranForm];
   Write[stmp,"MsqH      =      ", MsqHLL];
   Close[stmp], Print[" "]];

(* For the Transverse polarization put me=0, betae->1 *)
(* Recall Pol={PolLL,PolLM,PolLP,PolML,PolPL,PolPP,PolPM,PolMP,PolMM} *)
(*               1     2     3     4     5     6     7     8     9    *)

(****************************************************************************************)
\end{boxedverbatim}
\end{center}

\begin{center}
\begin{boxedverbatim}
MsqTL=Sum[MsqNuNu[[i]] + MsqGG[[i]] + MsqZZ[[i]] + \
          2*MsqNuG[[i]] + 2*MsqNuZ[[i]] + 2*MsqGZ[[i]] +\
          MsqHH[[i]] + 2*MsqNuH[[i]] + 2*MsqGH[[i]] + 2*MsqZH[[i]] \
          /. {me->0,betae->1} //Simplify,{i,2,5}] ;

MsqTL=Simplify[MsqTL /. steta->Sqrt[1-cteta^2]];
MsqTL=Simplify[MsqTL /. ct];
MsqTL=Simplify[MsqTL /. {me->0,betae->1}];
MsqTL=Simplify[MsqTL*g^4/4 /. weakangle] /. g^4->32 MW^4 GF^2;
MsqTL=Simplify[MsqTL /. beta2];

MsqTT=Sum[MsqNuNu[[i]] + MsqGG[[i]] + MsqZZ[[i]] + \
          2*MsqNuG[[i]] + 2*MsqNuZ[[i]] + 2*MsqGZ[[i]] +\
          MsqHH[[i]] + 2*MsqNuH[[i]] + 2*MsqGH[[i]] + 2*MsqZH[[i]] \
          /. {me->0,betae->1} //Simplify,{i,6,9}] ;

MsqTT=Simplify[MsqTT /. steta->Sqrt[1-cteta^2]];
MsqTT=Simplify[MsqTT /. ct];
MsqTT=Simplify[MsqTT /. {me->0,betae->1}];
MsqTT=Simplify[MsqTT*g^4/4 /. weakangle] /. g^4->32 MW^4 GF^2
MsqTT=Simplify[MsqTT /. beta2];

MsqTotal = Simplify[MsqLL + MsqTL + MsqTT]

(* Write Fortran Code *)

xs=1
If[xs==1,
   SetOptions[$Output, PageWidth -> 70];
   stmp=OpenWrite["ansTL.f",FormatType -> FortranForm];
   Write[stmp,"MsqTL = ", MsqTL];
   Close[stmp];
   stmp=OpenWrite["ansTT.f",FormatType -> FortranForm];
   Write[stmp,"MsqTT = ", MsqTT];
   Close[stmp];
   stmp=OpenWrite["ansTotal.f",FormatType -> FortranForm];
   Write[stmp,"MsqTotal = ", MsqTotal];
   Close[stmp], Print[" "]];
(****************************************************************************************)
\end{boxedverbatim}
\end{center}


\begin{thebibliography}{1}

\bibitem{quigg:1983}
C.~Quigg,
\newblock {\em Gauge theories of strong, weak and alectromagnetic interactions}
  (Addison-Wesley, New York, 1983).

\bibitem{romao:2016itc}
J.~C. Romão,
\newblock {\em Introdução à Teoria do Campo} (IST, 2016),
\newblock Available online at http://porthos.ist.utl.pt/ftp/textos/itc.pdf.

\bibitem{ctqft:2007}
J.~C. Romao,
\newblock http://porthos.ist.utl.pt/CTQFT/ .

\bibitem{udine:2012}
http://www.idpasc.lip.pt/ .

\end{thebibliography}
\end{document}